\documentclass[journal=jpcc,manuscript=article, layout=traditional]{achemso}
\setkeys{acs}{articletitle = true, etalmode=truncate, maxauthors = 10}
\usepackage{graphicx}
\usepackage{amsmath,amssymb,amsfonts,bm}
\usepackage{longtable}
\usepackage{siunitx}
\usepackage{times}
\usepackage{color}
\usepackage{soul}
\usepackage[normal]{subfigure}
\usepackage{gensymb}
\usepackage[table,xcdraw]{xcolor} 
\usepackage[utf8]{inputenc}
\usepackage{multirow}

\def\ha{(HA)$_2$PbI$_4$}
\def\haa{(HA)$_2$PbBr$_4$}
\def\ba{(BA)$_2$PbI$_4$}
\def\bna{(BNA)$_2$PbI$_4$}
\def\cm{cm$^{-1}$}

\title{Characterization of the Ammonium Bending Vibrations in Two-Dimensional Hybrid Lead-Halide Perovskites from Raman Spectroscopy and First-Principles Calculations}
\author{Sydney N. Lavan}
\affiliation{Department of Chemistry, Wayne State University, Detroit, MI, USA 48202}
\author{Adedayo M. Sanni}
\affiliation{Department of Chemistry, Wayne State University, Detroit, MI, USA 48202}
\author{Aaron S. Rury}
\affiliation{Department of Chemistry, Wayne State University, Detroit, MI, USA 48202}
\email{aaron.rury@wayne.edu}
\author{Zhen-Fei Liu}
\affiliation{Department of Chemistry, Wayne State University, Detroit, MI, USA 48202}
\email{zfliu@wayne.edu}
\begin{document}

\begin{abstract}
The facile synthesis and electronic properties of two-dimensional hybrid organic-inorganic perovskites (2D HOIPs) make these self-assembled systems an important class of energy materials. The basic building blocks of these materials include inorganic lattice frameworks that often consist of lead-halide octahedra and organic molecules possessing ammonium functional groups. Understanding the coupling between the inorganic and organic layers is key to unraveling how the electronic properties of 2D HOIPs relate to their structures. In this work, we leverage Raman spectroscopy measurements and first-principles calculations to characterize the Raman-active modes in four 2D HOIPs: hexylammonium lead iodide [\ha, HA = C$_6$H$_{13}$NH$_3^+$], hexylammonium lead bromide [\haa], butylammonium lead iodide [(BA)$_2$PbI$_4$, BA = C$_4$H$_9$NH$_3^+$], and benzylammonium lead iodide [(BNA)$_2$PbI$_4$, BNA = C$_6$H$_5$CH$_2$NH$_3^+$]. We focus on the 1400-1600 \cm~range where the Raman intensity of the molecular constituents is the strongest, and assign the major peaks observed in experiments as ammonium bending vibrations. We employ a combination of density functional perturbation theory based on the local density approximation and the frozen-phonon approach based on the vdw-DF-cx functional to find quantitative agreement between experimental and calculated Raman spectra. Furthermore, by comparing the vibrational spectra of isolated molecular cations with those near lead-halide clusters, we show how the inorganic lattice framework modulates the vibrational properties of the organic cations. We conclude that the properties of the Raman-active ammonium bending modes could effectively probe the local microscopic structure of the inorganic lattice framework in 2D HOIPs.
\end{abstract}

\section{Introduction}
Hybrid organic-inorganic perovskites (HOIPs) have been shown to be promising for optoelectronic applications \cite{doi:10.1063/1.453467,weber,Mitzi:1994aa,Mitzi1473,doi:10.1021/acs.chemrev.5b00715}. The prototypical and the most studied HOIP is the methylammonium lead halide with a completely interconnected three-dimensional (3D) inorganic structure, which attracted much attention due to their low cost \cite{C8CS00332G}, ease of synthesis \cite{C7TC00538E}, and excellent electrical and optical properties \cite{Correa-Baena739,Stranks:2015aa}. However, their relatively poor stability in air \cite{Gratzel:2017aa,Grancini:2017aa,C4TA04994B} under realistic operational conditions hinders future development of these 3D hybrid perovskites at the device level. In contrast, as an emerging class of materials, layered 2D HOIPs feature much better environmental stability \cite{doi:10.1002/anie.201406466,Tsai:2016aa} while maintaining the flexibility in tuning their electronic properties \cite{C9TC01325C,Ning:2017aa,C7TC02863F} as well as the desired performance in applications such as photovoltaics \cite{doi:10.1021/jacs.5b03796} and light emitting diodes \cite{doi:10.1021/acs.jpclett.8b01417}. In 2D HOIPs, the basic building blocks are inorganic layers that often consist of lead and halogen atoms and organic layers that often consist of long molecules with ammonium functional groups \cite{Abdel-Aal:2017aa}. The layered structure gives rise to distinct physics from 3D HOIPs, such as generally stronger exciton binding energies \cite{C9TC04292J}. Understanding these phenomena requires a complete characterization of the coupling between the inorganic and organic layers \cite{doi:10.1021/acs.jpcc.0c04573} and how the coupling depends on the atomistic details in the crystal structure \cite{doi:10.1021/acs.jpclett.5b01309}, which is currently missing for 2D HOIPs. 

Studies of the vibrational properties in HOIPs yield valuable information on the interplay between the inorganic framework and the organic molecule to provide physical insights into their dynamical interactions \cite{doi:10.1021/jz402589q,MAALEJ1997279,doi:10.1021/acs.jpcc.0c04573,doi:10.1021/acs.jpclett.5b01309,PLPM18}. For this purpose, both infrared (IR) and Raman spectroscopic studies have shown that the behavior of the organic cation depends on microscopic organic-inorganic interactions.\cite{doi:10.1021/acs.jpclett.5b01309,Ledinsky:2015aa,C6CP01723A,doi:10.1021/acs.jpcc.0c04573}. For 3D HOIPs, several studies indicated the facets of the coupled organic-inorganic structural dynamics\cite{Bakulin:2015aa}, including the role of dynamic disorder \cite{Frost:2016aa}, polar fluctuations, \cite{PhysRevLett.118.136001} and static dielectric environments\cite{doi:10.1021/acs.jpclett.5b01309,Ivanovska:2016aa}. Additionally, Ref. \citenum{Park:2017aa} focused on the dynamic structure-property relationships and found that the Raman activity of the methylammonium molecule is very sensitive to structural distortions of the inorganic framework. For 2D HOIPs, using temperature- and polarization-dependent measurements, Ref. \citenum{Dhanabalan_2020} found strong evidence of anisotropy in the vibrational bands in phenethylammonium lead bromide, and attributed the anisotropy to the orientation of the phenethyl ring with respect to the inorganic lattice framework. Given the importance played by the ammonium functional group in the coupling between the inorganic and organic layers in 2D HOIPs, carefully characterizing its vibrational properties is central to understanding the interactions between the inorganic lead-halide framework and the organic spacer cations.

In this work, we employ a combination of experimental Raman spectroscopy and first-principles calculations to study the vibrational properties of four 2D HOIPs: hexylammonium lead iodide [\ha, where HA = C$_6$H$_{13}$NH$_3^+$], hexylammonium lead bromide [\haa], butylammonium lead iodide [(BA)$_2$PbI$_4$, where BA = C$_4$H$_9$NH$_3^+$], and benzylammonium lead iodide [(BNA)$_2$PbI$_4$, where BNA = C$_6$H$_5$CH$_2$NH$_3^+$]. Calculations based on the density functional perturbation theory (DFPT)\cite{RevModPhys.73.515} provide orientation-dependent contributions to the Raman intensity, which are compared to experimental Raman spectra taken under different incident and scattered field polarization configurations. A frozen-phonon approach based on DFPT normal-mode eigenvectors is used to correct the intrinsic errors of the underlying density functional employed in the DFPT calculations. The quantitative agreement between experiment and computation enables us to unambiguously assign experimental peaks with specific normal modes of the crystal. We find that three groups of ammonium bending modes dominate the Raman intensity between 1400 \cm~and 1600 \cm~for all materials studied. By comparing DFPT results of the crystal and those of the organic molecular cation placed next to a fragment of the inorganic framework, we conclude that the crystal effect on the Raman properties of the organic molecule is local, which could be used as an effective probe for the atomistic detail and possible defects in the lead-halide framework.

This paper is structured as follows. We first describe our experimental and computational methods: preparation of the materials, Raman scattering measurements, and first-principles calculations of the structural and vibrational properties. Then we make a detailed comparison between measured Raman spectra and calculated Raman intensities. After that, we analyze the effects of different functionals in order to achieve quantitative agreement between experiment and computation. Lastly, we focus on the ammonium bending modes and reveal the crystal effects in modulating the frequencies of such modes compared to the case of an isolated molecule, before we draw a brief conclusion. For the conciseness of this article, we discuss the results for the \ha~crystal in detail in the main text, and list the results for other materials in the Supporting Information.
 
 \section{Methods}
\subsection{Preparation of 2D Hybrid Organic-Inorganic Perovskites}
The starting chemicals were purchased from Sigma-Aldrich and synthesized without further purification. We prepared our perovskite crystal structures using the methods previously reported. \cite{doi:10.1021/acs.jpcc.0c04573,doi:10.1021/acs.jpclett.9b00743,doi:10.1021/cm9505097,doi:10.1021/acs.chemmater.6b00809}

Specifically, we synthesized (HA)$_2$PbI$_4$ by dissolving 74 mg of PbI$_2$ in 3 mL of HI in a beaker and transferred gently into a test tube using a syringe. To the test tube containing the PbI$_2$ in HI solution, we gently added 3 mL of methanol. Due to the difference in densities, an interface was formed between the solvents. we then gently added 1 mL hexylamine to the tube using a syringe. We covered the test tube with thin foil. Orange plate-like crystals started forming within 24 hrs. We let more crystals form until 5 days before filtering the crystals under suction. The orange crystals were washed with cold diethyl ether, dried in the oven at 60 $^{\circ}$C for 24 hrs.

\haa~was synthesized by dissolving 58.5 mg of PbBr$_2$ in 3 mL of 48\% HBr in a beaker and transferred gently into a test tube using a syringe. To the test tube containing the PbBr$_2$ in HBr solution, 3 mL of methanol was gently added using a syringe. Hexylamine (1 mL) was then gently added to the test tube and the test tube was covered with thin foil. White plate-like crystals started forming within 24 hrs. We let more crystals form until 5 days before filtering the crystals under suction. The white crystals were washed with cold diethyl ether, dried in the oven at 60 $^{\circ}$C for 24 hrs.

To synthesize (BA)$_2$PbI$_4$, we dissolved and heat to 130 $^{\circ}$C, 1.16 mmol lead iodide in 2 mL HI solution in one beaker. We neutralized 2.32 mmol butylamine with 3 mL HI in a second beaker. The neutralized butylamine solution was then added to the lead iodide solution while heating and stirring for approximately 20 minutes under a dry nitrogen atmosphere. The solution was then cooled to room temperature and orange crystals formed. The crystallization was completed in an ice bath for 2 hrs and filtered crystals were washed and dried in a vacuum oven at 60 $^{\circ}$C for 24 hrs.

We synthesized (BNA)$_2$PbI$_4$ by following similar method used in synthesizing (HA)$_2$PbI$_4$. We dissolved 74 mg of PbI$_2$ in 3 mL of 57\% HI in a beaker, followed by transferring the solution gently into a test tube using a syringe. To the test tube containing the PbI$_2$ in HI solution, 3 mL of methanol was gently placed on the lead iodide solution. We gently added 0.5 mL benzylamine to the test tube and covered the test tube with thin foil. Orange plate-like crystals started forming within 24 hrs. We let more crystals form until 4 days before filtering the crystals by suction. The orange crystals were washed with cold diethyl ether and dried in the oven at 60 $^{\circ}$C for 24 hrs.

\subsection{Raman Scattering Measurements}
We collected our experimental Raman scattering spectra with a Horiba XploRA PLUS Raman micro-spectrometer. The excitation laser light was set at 785 nm and dispersed in a 0.3 m monochromator using a 1800 g/mm diffraction grating. Our instrument is setup in a 180\si{\degree} backscattering arrangement and allows measurements dependent on both the incidence and scattered electric field polarizations. We choose the direction of the polarization which is either parallel (VV, vertical) or perpendicular (HV, horizontal) to the polarization of the excitation laser. A Linkam THMS600 liquid-nitrogen temperature probe controlled the temperature of each sample with a precision of $\pm$0.5 K. We allowed the sample to settle for approximately 5 min at each desired temperature before acquiring the spectra.
  
\subsection{First-Principles Calculations} 
We employ first-principles calculations to assign experimental peaks in Raman spectra as specific vibrational modes and to understand the effect of the crystal structure in modulating the vibrational properties of the organic molecular cation. We first optimize the structures of each material using density functional theory (DFT)\cite{PhysRev.136.B864,PhysRev.140.A1133}, and subsequently calculate their vibrational properties based on DFPT \cite{RevModPhys.73.515} as implemented in the \texttt{Quantum ESPRESSO} \cite{Giannozzi_2017} package. The Perdew-Zunger form of local density approximation (LDA)\cite{PhysRevB.23.5048,PhysRevLett.45.566} was used for all geometry optimization and DFPT calculations. The pseudopotentials are adapted from Refs. \citenum{VANSETTEN201839,PhysRevB.88.085117}. 

For the geometry optimization of \ha, \ba, and \bna, we use the crystal structure resolved from X-ray diffraction experiments\cite{Groom:bm5086} as the initial guess. During the optimization, all lattice parameters and atomic coordinates are relaxed while keeping the lattice symmetry to be the experimentally determined one [monoclinic for \ha~and orthorhombic for \ba~and \bna]. For \haa, we have not found experimentally determined lattice structure, so we assume it has the same space group as \ha~and use the \ha~structure with all iodine atoms replaced by bromine as the initial guess. For \ha~and \haa, we have checked that a $\mathbf{k}$-mesh of 5$\times$5$\times$3 and a kinetic energy cutoff of 100 Ry lead to a convergence in total energies of the unit cell within 5 meV/atom, and so we use a cutoff of 130 Ry in the variable-cell relaxation to mitigate the Pulay stress. For \ba~and \bna, a $\mathbf{k}$-mesh of 5$\times$5$\times$2 and a cutoff of 130 Ry were used. The optimization is considered complete when all forces are below 0.05 eV/\AA~and the pressure of the cell is below 0.5 kbar.  

After the optimized structures are obtained, we proceed with the self-consistent calculation of the electronic structure using a kinetic energy cutoff of 100 Ry and the same $\mathbf{k}$-mesh as above, before the vibrational calculations using DFPT. Since the Raman experiments only measure $\Gamma$-centered modes \cite{Giustino}, we constrain our DFPT calculations of the phonon normal modes and Raman activities \cite{PhysRevLett.90.036401} to the $\Gamma$ point only. The crystal acoustic sum rule is applied in the diagonalization of the dynamical matrix, and we have checked that it correctly yields three zero-frequency modes. Using a slightly tailored version of the code, we output each component of the Raman activities, $\left|\partial \alpha_{ij}/\partial Q_\nu\right|^2$, where $\alpha_{ij}$ is the polarizability tensor with $i, j=x, y,$ or $z$ and $Q_\nu$ denotes the normal-mode coordinate. 

\section{Results and Discussions}
\subsection{Crystal Structure}

\begin{figure}[h]
\centering
\includegraphics[width=0.6\columnwidth]{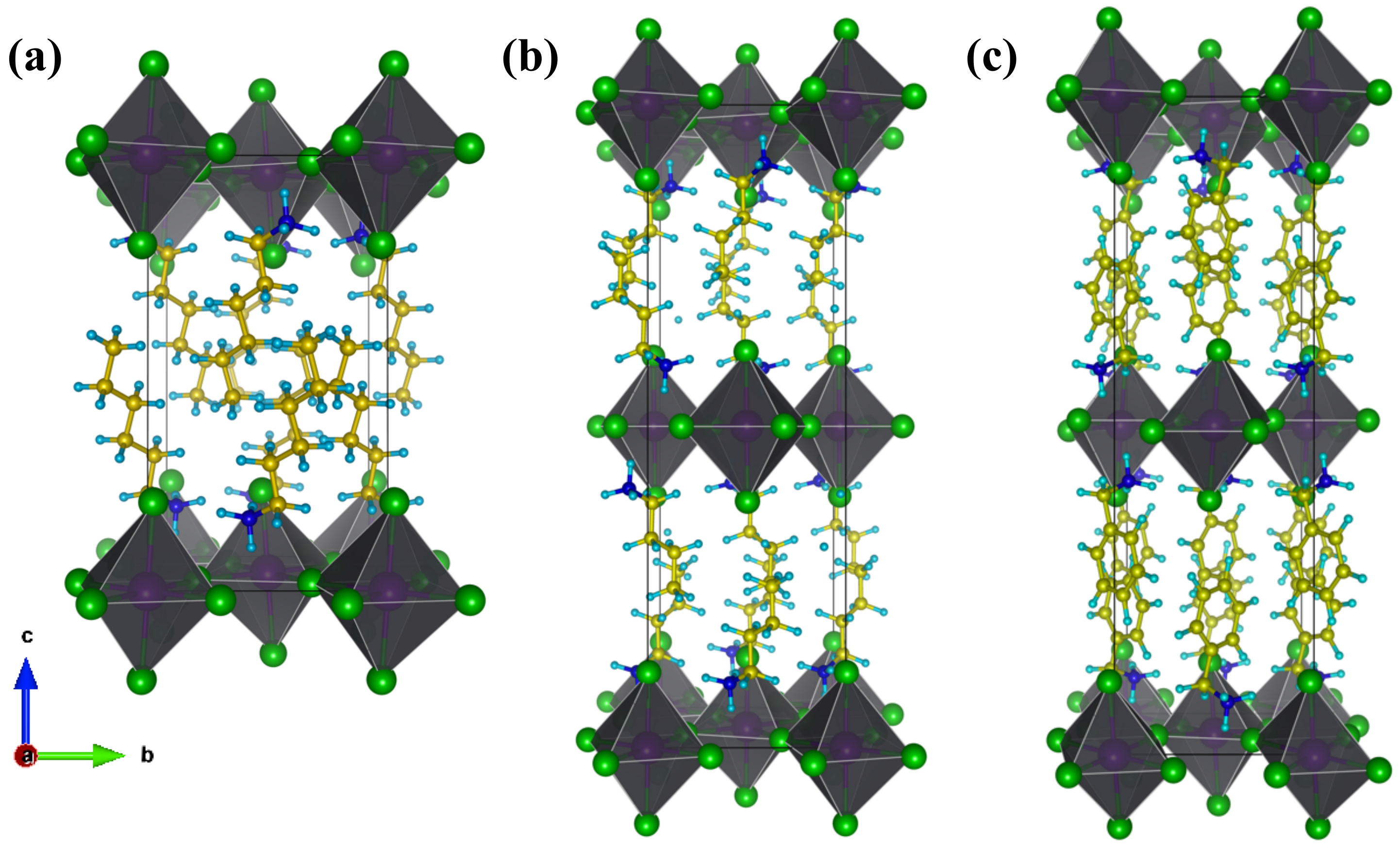}
 \caption{The optimized crystal structure of (a) \ha, (b) \ba, and (c) \bna. The corner-sharing PbI$_6$ octahedra are shown. The structure of \haa~is similar to (a) with iodide atoms replaced by bromide. Color code: C - yellow; H - cyan; N - blue; Pb - magenta; I - green. This figure is rendered using VESTA \cite{Momma:ko5060}.}
 \label{fig:crys}
 \end{figure}

Fig. \ref{fig:crys}(a) shows the optimized unit cell of \ha, with a base-centered monoclinic crystal structure. Two lead atoms occupy the corners and the face centers of the (001) plane. The lead and iodine atoms form corner-sharing octahedra in the (001) plane, with four HA$^+$ cations filled between two adjacent layers of Pb-I octahedra. The structure of \haa~is very similar to that of \ha, with bromine atoms replacing iodine atoms and slightly different lattice parameters. Fig. \ref{fig:crys}(b)(c) show the optimized unit cells of \ba~and \bna, with a face-centered orthorhombic crystal structure. One lead atom occupies the corners and three lead atoms occupy the face centers. The lead and iodine atoms form corner-sharing octahedra in both the (001) and the (002) planes. Four organic spacer cations are filled between each two adjacent layers of Pb-I octahedra and there are overall eight organic cations in the unit cell.

\begin{table}[h]
\begin{tabular}{cccc}
\hline\hline
2D HOIP &  Symmetry &  Experiment &  DFT Calculation \\ 
\hline
(HA)$_2$PbI$_4$ &
  \begin{tabular}[c]{@{}c@{}}Monoclinic\\ \\ P2$_1$/a\end{tabular} &
  \begin{tabular}[c]{@{}l@{}}$a = 8.643$ \AA \\ $b = 8.845$ \AA \\ $c = 16.042$ \AA \\ $\alpha = \gamma = 90^{\circ}$ \\ $\beta = 91.985^{\circ}$\\ $V = 1226.40$ \AA$^3$\end{tabular} &
  \begin{tabular}[c]{@{}l@{}}$a = 8.343$ \AA \\ $b = 8.475$ \AA \\ $c = 15.397$ \AA \\ $\alpha = \gamma = 90^{\circ}$\\ $\beta = 90.708^{\circ}$\\ $V = 1088.77$ \AA$^3$\end{tabular} \\ \hline
(HA)$_2$PbBr$_4$ &
  \begin{tabular}[c]{@{}c@{}}Monoclinic\\ \\ P2$_1$/a\end{tabular} &
  - &
  \begin{tabular}[c]{@{}l@{}}$a= 7.936$ \AA \\ $b= 8.042$ \AA \\ $c= 15.301$ \AA \\ $\alpha=\gamma=90^{\circ}$ \\ $\beta= 90.328^{\circ}$\\ $V = 976.77$ \AA$^3$ \end{tabular} \\ 
  \hline
 (BA)$_2$PbI$_4$ &
  \begin{tabular}[c]{@{}c@{}}Orthorhombic\\ \\ Pbca\end{tabular} &
  \begin{tabular}[c]{@{}l@{}}$a = 8.876$ \AA \\ $b = 8.693$ \AA \\ $c = 27.601$ \AA \\ $\alpha = \beta = \gamma = 90^{\circ}$\\ $V = 2129.67$ \AA$^3$\end{tabular} 
  &
  \begin{tabular}[c]{@{}l@{}}$a = 8.396$ \AA \\ $b = 8.329$ \AA \\ $c = 27.033$ \AA \\ $\alpha = \beta = \gamma = 90^{\circ}$\\ $V = 1890.44$ \AA$^3$\end{tabular} \\ 
  \hline
(BNA)$_2$PbI$_4$ &
  \begin{tabular}[c]{@{}c@{}}Orthorhombic\\ \\ Pbca\end{tabular} &
  \begin{tabular}[c]{@{}l@{}}$a = 9.111$ \AA \\ $b = 8.634$ \AA \\ $c = 28.408$ \AA \\ $\alpha = \beta = \gamma = 90^{\circ}$\\ $V= 2234.67$ \AA$^3$\end{tabular}
   &
  \begin{tabular}[c]{@{}l@{}}$a = 8.870$ \AA \\ $b = 8.330$ \AA \\ $c = 27.490$ \AA \\ $\alpha = \beta = \gamma = 90^{\circ}$\\ $V = 2031.26$ \AA$^3$\end{tabular} \\
 \hline\hline
\end{tabular}
\caption{A comparison of crystallographic parameters between experiment and DFT calculations.}
\label{tab:crys}
\end{table}

In Table \ref{tab:crys}, we compare the crystallographic parameters from experiment and our DFT calculations. The experimental crystal structures are obtained from X-ray diffraction measurements after the synthesis and growth of the material.\cite{doi:10.1021/acs.jpclett.9b00743} We have not been able to resolve the crystal structure of \haa~ourselves or find reported experimental values in the literature. Our LDA-based calculations underestimate lattice parameters by about 3\%-5\% and the crystal volume by about 10\%, which is consistent with the typical performance of LDA in bond lengths and lattice parameters. \cite{PhysRevB.89.064305}

\subsection{Symmetry Analysis and Raman Spectra}
For \ha~and \haa, the crystal structure is monoclinic with space group P2$_1$/a and point group C$_{\rm 2h}$. In total, there are 102 atoms in the unit cell, giving rise to 306 normal modes. These modes admit the following symmetry representation: $\Gamma= 78\mbox{A}_{\rm u} + 78\mbox{B}_{\rm u} + 75\mbox{A}_{\rm g} + 75\mbox{B}_{\rm g}$. All A$_{\rm u}$ and B$_{\rm u}$ modes are IR active and all A$_{\rm g}$ and B$_{\rm g}$ modes are Raman active, satisfying the rule of mutual exclusion given the presence of the centrosymmetry in P2$_1$/a. Our DFPT results align well with this symmetry analysis. The general Raman tensors for C$_{\rm 2h}$ are: \cite{Lines:1977aa}
\begin{equation}
\alpha(\mbox{A}_{\rm g})=\begin{pmatrix}
        a & 0 & d \\
        0 & b & 0 \\
        d & 0 & c 
    \end{pmatrix};
\hspace{0.3in} 
\alpha(\mbox{B}_{\rm g})=\begin{pmatrix}
        0 & e & 0 \\
        e & 0 & f \\
        0 & f & 0 
    \end{pmatrix}.
    \label{eq:ha}
\end{equation}

For \ba~and \bna, the crystal structure is orthorhombic with space group Pbca and point group D$_{\rm 2h}$. For \ba, there are 156 atoms in the unit cell and 468 normal modes overall, admitting the following symmetry representation: $\Gamma = 60 \mbox{A}_{\rm u} +  60 \mbox{B}_{\rm 1u} +  60 \mbox{B}_{\rm 2u} +  60 \mbox{B}_{\rm 3u} +  57 \mbox{A}_{\rm g} + 57 \mbox{B}_{\rm 1g} + 57  \mbox{B}_{\rm 2g} +  57 \mbox{B}_{\rm 3g}$. For \bna, there are 164 atoms in the unit cell and 492 normal modes overall, admitting the following symmetry representation: $\Gamma = 63 \mbox{A}_{\rm u} + 63 \mbox{B}_{\rm 1u} + 63 \mbox{B}_{\rm 2u} + 63 \mbox{B}_{\rm 3u} + 60 \mbox{A}_{\rm g} + 60 \mbox{B}_{\rm 1g} + 60 \mbox{B}_{\rm 2g} + 60 \mbox{B}_{\rm 3g}$. Of these, the B$_{\rm 1u}$, B$_{\rm 2u}$, and B$_{\rm 3u}$ modes are IR active, all the ``g'' modes are Raman active, and the A$_{\rm u}$ modes are neither IR or Raman active, consistent with the character table of D$_{\rm 2h}$. The general Raman tensors for D$_{\rm 2h}$ are \cite{Lines:1977aa}:
\begin{equation}
\alpha(\mbox{A}_{\rm g})=\begin{pmatrix}
        a & 0 & 0 \\
        0 & b & 0 \\
        0 & 0 & c 
    \end{pmatrix};
\hspace{0.1in}
\alpha(\mbox{B}_{\rm 1g})=\begin{pmatrix}
        0 & d & 0 \\
        d & 0 & 0\\
        0 & 0 & 0 
    \end{pmatrix};
\hspace{0.1in}
\alpha(\mbox{B}_{\rm 2g})=\begin{pmatrix}
        0 & 0 & e \\
        0 & 0 & 0 \\
        e & 0 & 0 
    \end{pmatrix};
\hspace{0.1in}
\alpha(\mbox{B}_{\rm 3g})=\begin{pmatrix}
        0 & 0 & 0 \\
        0 & 0 & f \\
        0 & f & 0 
    \end{pmatrix}.
    \label{eq:ba}
\end{equation}

Using Porto's notation, our back-scattering Raman measurement in the VV polarization has the $z(xx)\bar{z}$ or $z(yy)\bar{z}$ geometry, and our measurement in the HV polarization has the $z(xy)\bar{z}$ geometry. Here, $x$, $y$, and $z$ are the [100], [010], and [001] crystallographic directions. Therefore, the VV measurement probes the $\left|\partial \alpha_{xx}/\partial Q_\nu\right|^2$ and the $\left|\partial \alpha_{yy}/\partial Q_\nu\right|^2$ contributions to the Raman intensity, corresponding to the matrix elements $a$ and $b$ in the A$_{\rm g}$ Raman tensors in both Eqs. \eqref{eq:ha} and \eqref{eq:ba}. The HV measurement probes the $\left|\partial \alpha_{xy}/\partial Q_\nu\right|^2$ contribution to the Raman intensity, corresponding to the matrix element $e$ in the B$_{\rm g}$ Raman tensor in Eq. \eqref{eq:ha} and the matrix element $d$ in the B$_{\rm 1g}$ Raman tensor in Eq. \eqref{eq:ba}. 

\begin{figure}[htbp]
\centering
\includegraphics[width=0.6\columnwidth]{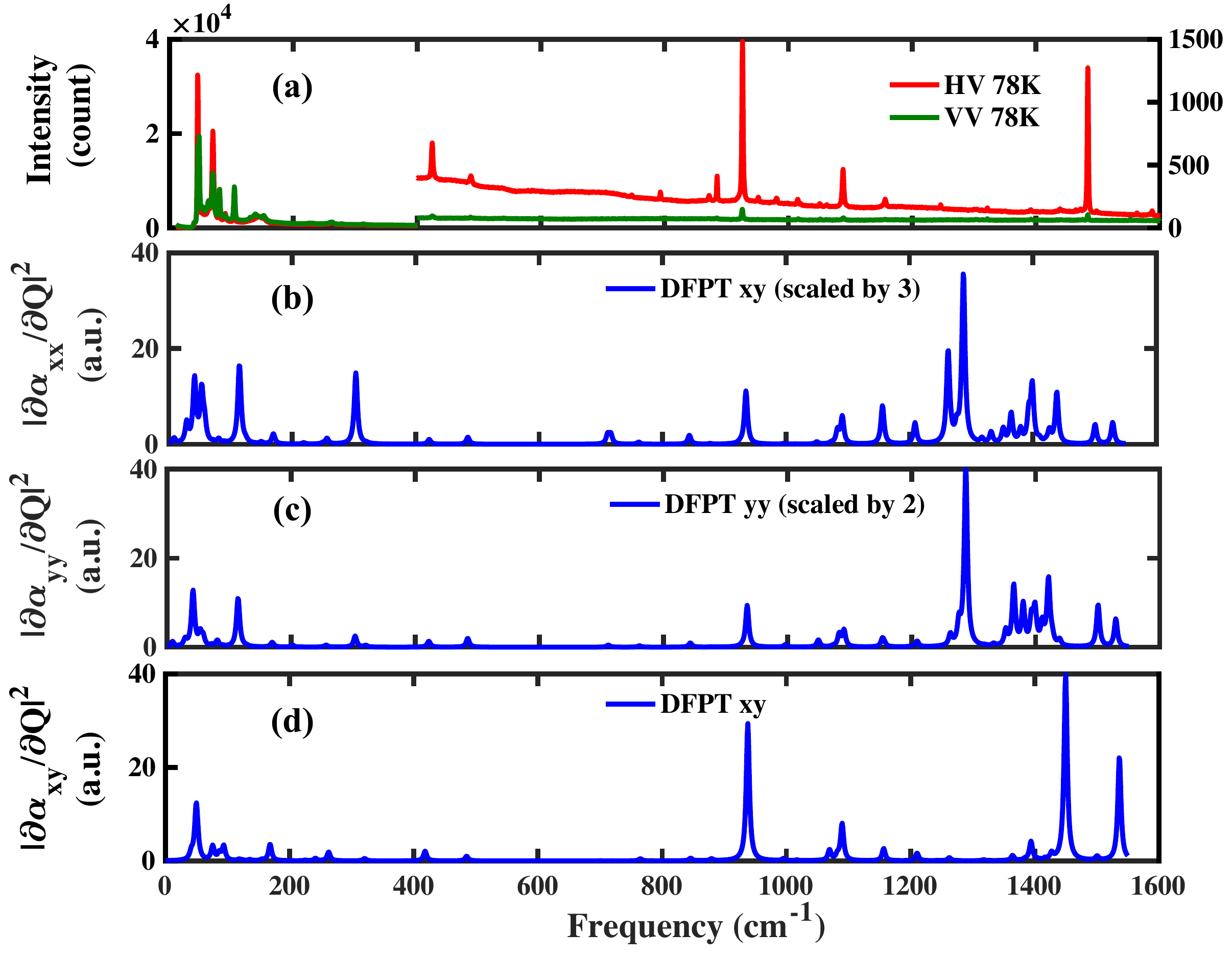}
 \caption{Comparison between measured and calculated Raman spectra for \ha. (a) The experimentally measured Raman intensities, in HV (red) and VV (green) polarizations at 78 K. Due to the large intensities of the low-frequency modes, we use different scales in the vertical axis below (left axis) and above (right axis) 400 \cm. (b) The $xx$ contribution to the calculated Raman intensity, scaled up by 3 times. (c) The $yy$ contribution to the calculated Raman intensity, scaled up by 2 times. (d) The $xy$ contribution to the calculated Raman intensity.}
 \label{fig:compare}
 \end{figure}

For \ha, we compare the experimentally measured Raman intensities in both VV and HV polarizations and calculated $xx$, $yy$, and $xy$ contributions to the Raman intensities in Figure \ref{fig:compare}. In this work, for all calculated Raman intensities, we apply a Lorentzian broadening to each mode with a 3 \cm~half width at half maximum in the figures, and use atomic units (a.u.) throughout, i.e., Bohr$^4$/amu with amu the atomic mass unit. One can see that the VV (HV) qualitatively agrees with the $xx$ and $yy$ ($xy$) contributions to the Raman intensity. Due to the large intensities of the low-frequency modes seen in experiment, we use different scales in the vertical axis below and above 400 \cm~in Figure \ref{fig:compare}(a). We have also scaled Figures \ref{fig:compare}(b)(c)(d) differently in order to use a consistent range in their vertical axes for a clear comparison. Since the backscattering geometry in our experiment does not directly probe the $xz$, $yz$, or $zz$ contributions to the Raman intensity, we do not include those components in Figure \ref{fig:compare}. 

For all materials, the nature of all phonon modes can be categorized into the following: the modes below 500 \cm~are combinations of the translations/vibrations of the lead-halogen framework and the librations/hindered rotations of the organic molecules. The modes between 500 \cm~and 1650 \cm~are the bending modes of the organic molecule. Specifically, all the bending modes of the ammonium group are between 1400 \cm~and 1600 \cm, which we focus on in this paper. No modes are within the 1650-2850 \cm~range. Modes above 2850 \cm~are the H-X (X = C, N) bond stretching modes of the organic molecule, which we cannot detect with our experimental apparatus. 

\subsection{Mode Assignments between 1400 \cm~and 1600 \cm}

We note from Figure \ref{fig:compare}(a) that the strongest Raman intensity appears in two frequency ranges: 800-1000 \cm~and 1400-1600 \cm. The former mainly consists of the bending modes of the carbon backbone of the organic molecule sandwiched between the inorganic layers. Given the complexity of the molecular backbone structure and the variety of organic molecules used in 2D HOIPs, such modes are unlikely representative across a broad range of organic molecules in various 2D HOIPs. Furthermore, since the molecular backbone is relatively far away from the lead-halogen framework, their bending modes could not sensitively reflect the effect of the lead-halogen framework. On the other hand, the latter region, 1400-1600 \cm, mainly consists of the bending modes of the -NH$_3$ group in the organic molecule, which is both ubiquitous in many 2D HOIPs and close to the inorganic layer of the crystal. Since our goal is to study the coupling between the inorganic and organic layers, we focus on the 1400-1600 \cm~in this work.

\begin{figure}[htbp]
\centering
\includegraphics[width=0.6\columnwidth]{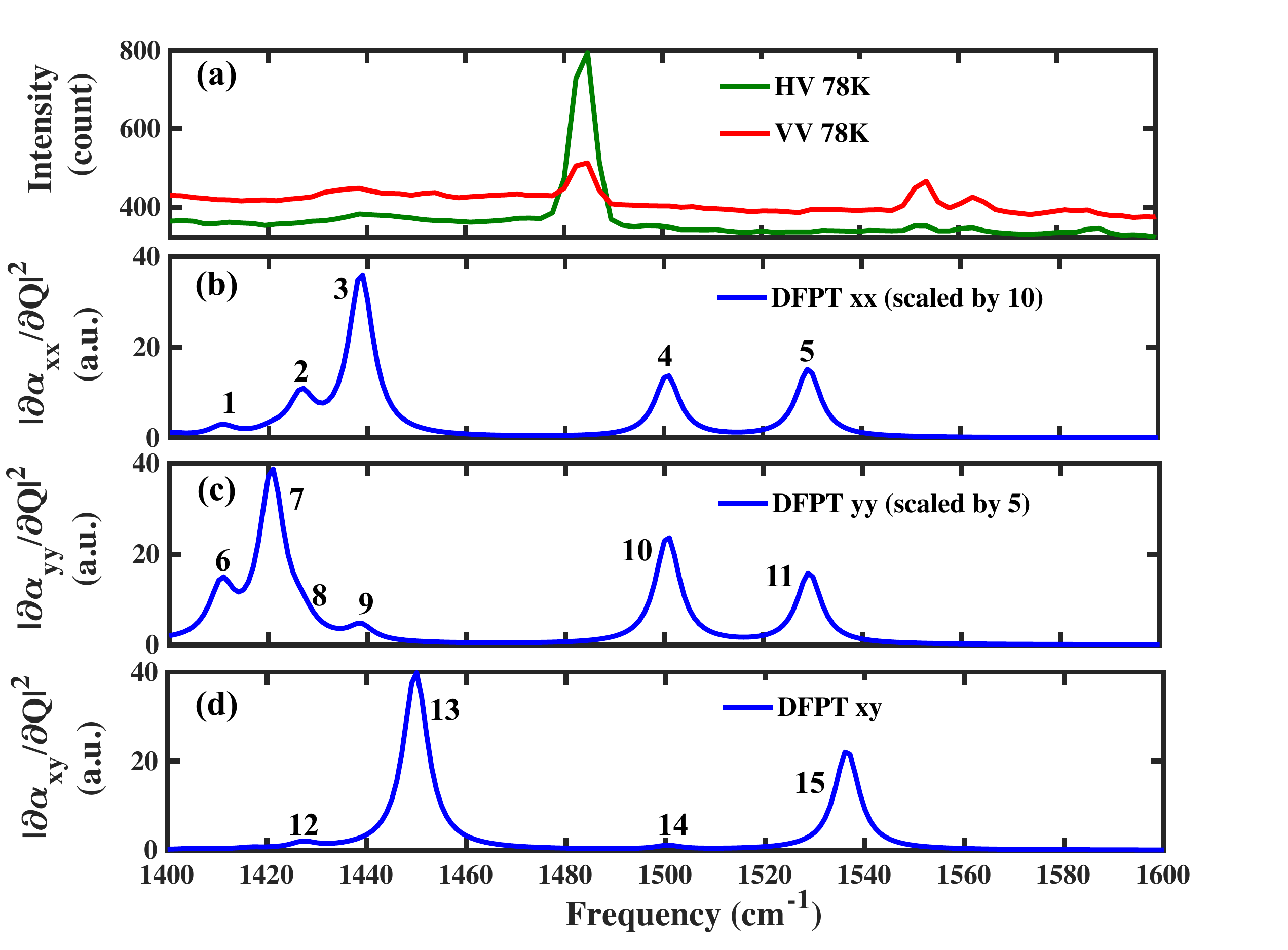}
 \caption{The Raman activities of \ha~between 1400 \cm~and 1600 \cm. (a) Measured Raman intensities in HV (green) and VV (red) polarizations. (b) The $xx$ contribution to the calculated Raman intensity, scaled up by 10 times. (c) The $yy$ contribution to the calculated Raman intensity, scaled up by 5 times. (d) The $xy$ contribution to the calculated Raman intensity. Peak numbers are shown to facilitate references in Table \ref{tab:ha} and Figure \ref{fig:hamodes}.}
 \label{fig:ha}
 \end{figure}

Figure \ref{fig:ha} shows the measured and calculated Raman activities from DFPT, for \ha~between 1400 \cm~and 1600 \cm. Figure \ref{fig:ha}(a) includes the experimental Raman spectra of both polarizations VV (red) and HV (green) at 78 K. We observe a major peak at 1484.7 \cm, and two minor ones at 1553.5 \cm~and 1562.9 \cm, respectively. In Figures \ref{fig:ha}(b)(c)(d), we show several contributions to the calculated Raman intensity: $xx$ in (b), $yy$ in (c), and $xy$ in (d). To make the range in the vertical axes the same for (b)(c)(d), we have scaled up the values in (b) by a factor of 10 and (c) by a factor of 5. From our symmetry analysis above, Figures \ref{fig:ha}(b)(c) correspond to the VV measurement and Figure \ref{fig:ha}(d) corresponds to the HV measurement. 

In the 1400-1600 cm$^{-1}$ range, there are overall 28 phonon modes with 14 of them Raman active. We recognize that the 28 modes can be divided into 7 groups with 4 modes in each group having similar frequencies, consistent with the fact that there are 4 molecules in the unit cell. Among the 4 modes in each group, two are Raman-active, one in A$_{\rm g}$ and one in B$_{\rm g}$, as we show in Table \ref{tab:ha}. For each mode, we show the values of the six orientation-dependent contributions to the Raman intensity. Based on Eq. \eqref{eq:ha}, the $xx$, $yy$, $zz$, and $xz$ components have A$_{\rm g}$ symmetry and the $xy$ and $yz$ components have B$_{\rm g}$ symmetry. To facilitate the discussion, we label each peak in Figures \ref{fig:ha}(b)(c)(d) from 1 to 15, which are matched to each mode in Table \ref{tab:ha}.

\begin{table}[htbp]
\begin{tabular}{cccccccc}
\hline\hline
\multicolumn{2}{c}{ } &  \multicolumn{4}{c}{A$_{\rm g}$} &  \multicolumn{2}{c}{B$_{\rm g}$} \\
Peak number in Figure \ref{fig:ha} & Frequency (\cm) & $xx$ & $yy$ & $zz$ & $xz$ & $xy$ & $yz$ \\
\hline\hline
-    & 1400.0 A$_{\rm g}$ & 0.07 & 0.05 & 0.00 & 2.85 & -     & -    \\
-    & 1403.8 B$_{\rm g}$ & -    & -    & -    & -    & 0.13  & 7.44 \\ \hline
-    & 1408.2 B$_{\rm g}$ & -    & -    & -    & -    & 0.04  & 7.41 \\
1, 6  & 1410.6 A$_{\rm g}$ & 0.22 & 2.35 & 0.25 & 1.36 & -     & -    \\ \hline
-    & 1416.8 B$_{\rm g}$ & -    & -    & -    & -    & 0.30  & 0.23 \\
7    & 1420.7 A$_{\rm g}$ & 0.09 & 7.44 & 0.17 & 0.00 & -     & -    \\ \hline
2, 8  & 1426.6 A$_{\rm g}$ & 0.86 & 0.81 & 3.32 & 0.42 & -     & -    \\
12   & 1427.3 B$_{\rm g}$ & -    & -    & -    & -    & 1.35  & 1.23 \\ \hline
3, 9  & 1438.7 A$_{\rm g}$ & 3.58 & 0.67 & 0.18 & 0.27 & -     & -    \\
13   & 1449.8 B$_{\rm g}$ & -    & -    & -    & -    & 39.98 & 0.02 \\ \hline
14   & 1500.5 B$_{\rm g}$ & -    & -    & -    & -    & 0.84  & 1.97 \\
4, 10 & 1500.6 A$_{\rm g}$ & 1.36 & 4.74 & 0.02 & 1.81 & -     & -    \\ \hline
5, 11 & 1529.2 A$_{\rm g}$ & 1.50 & 3.13 & 0.04 & 0.04 & -     & -    \\
15   & 1536.4 B$_{\rm g}$ & -    & -    & -    & -    & 22.34 & 0.04 \\ \hline\hline
\end{tabular}
\caption{All Raman active modes of \ha~between 1400 \cm~and 1600 \cm. The corresponding peak numbers in Figure \ref{fig:ha}, and the values of the six orientation-dependent contributions to the Raman intensity are shown next to each mode. All Raman intensities are in atomic units.}
\label{tab:ha}
\end{table}

\begin{figure}[h]
\includegraphics[width=\columnwidth]{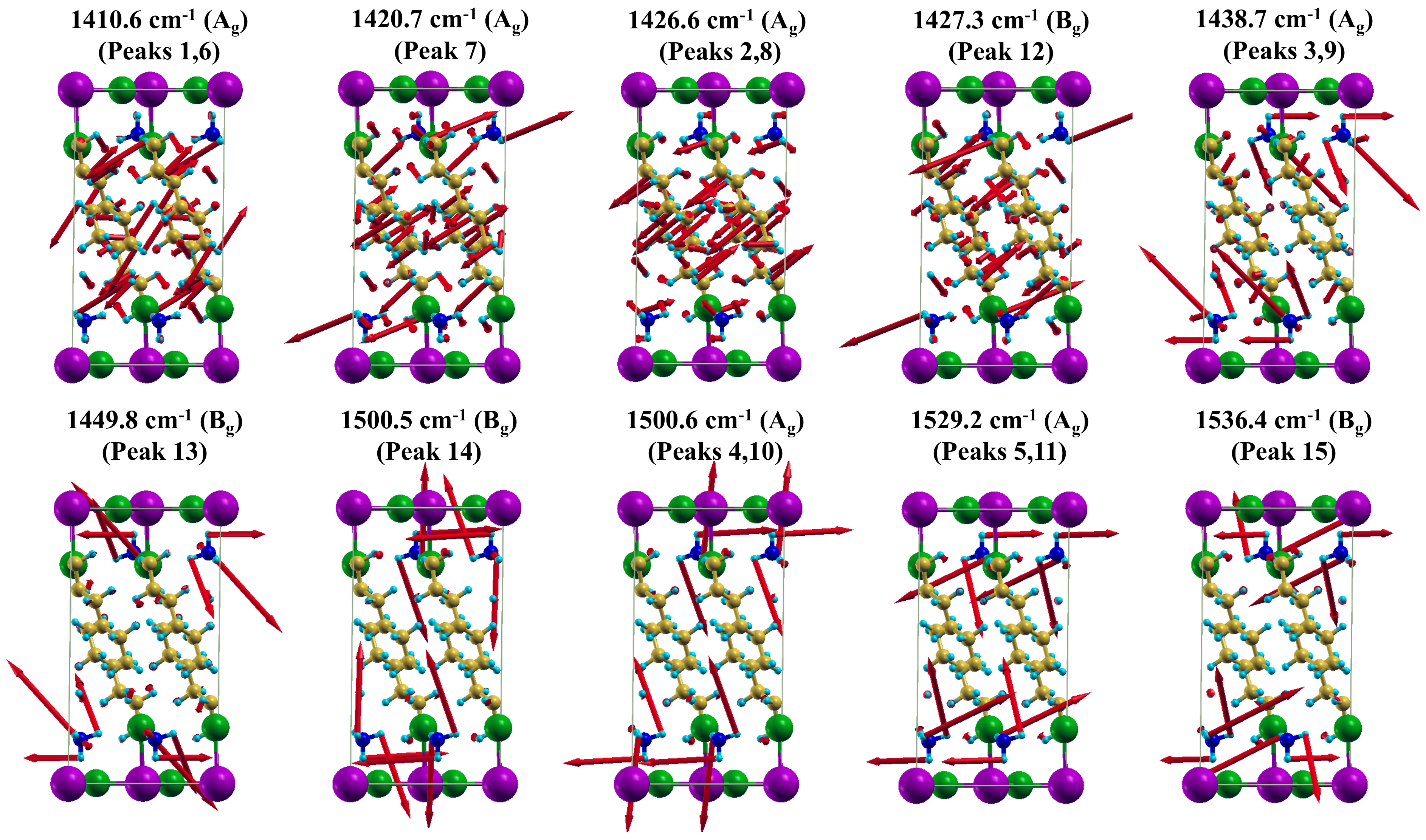}
\caption{A schematic view of all \ha~normal-mode eigenvectors that are included in Figures \ref{fig:ha}(b)(c)(d). The red arrows depict the direction and magnitude of the vibrational movements for each atom. The peak number in Figure \ref{fig:ha}, frequency, and symmetry for each mode are also shown. This figure is rendered using XCrySDen \cite{KOKALJ1999176}.}
\label{fig:hamodes}
\end{figure}

Figure \ref{fig:hamodes} shows the normal-mode eigenvectors of all modes that are included in Figures \ref{fig:ha}(b)(c)(d). One can see that the two modes in each group in Table \ref{tab:ha} consist of approximately the same vibrational patterns in all four molecules, and are only different in their symmetry over the unit cell of the crystal. We note that there are six modes that are primarily localized on the -NH$_3$ group and correspond to the ammonium bending modes, which are the last three rows in Table \ref{tab:ha} and the last six modes in Figure \ref{fig:hamodes}. The two strongest intensities in Figure \ref{fig:ha}(d) - one at 1449.8 \cm~(peak 13) and one at 1536.4 \cm~(peak 15) - are among these six ammonium bending modes, which suggests that such -NH$_3$ bending modes are useful experimental probes in studies of the vibrational properties of the crystal\cite{doi:10.1021/acs.jpcc.0c04573}. 

\begin{figure}[htbp]
\centering
\includegraphics[width=0.6\columnwidth]{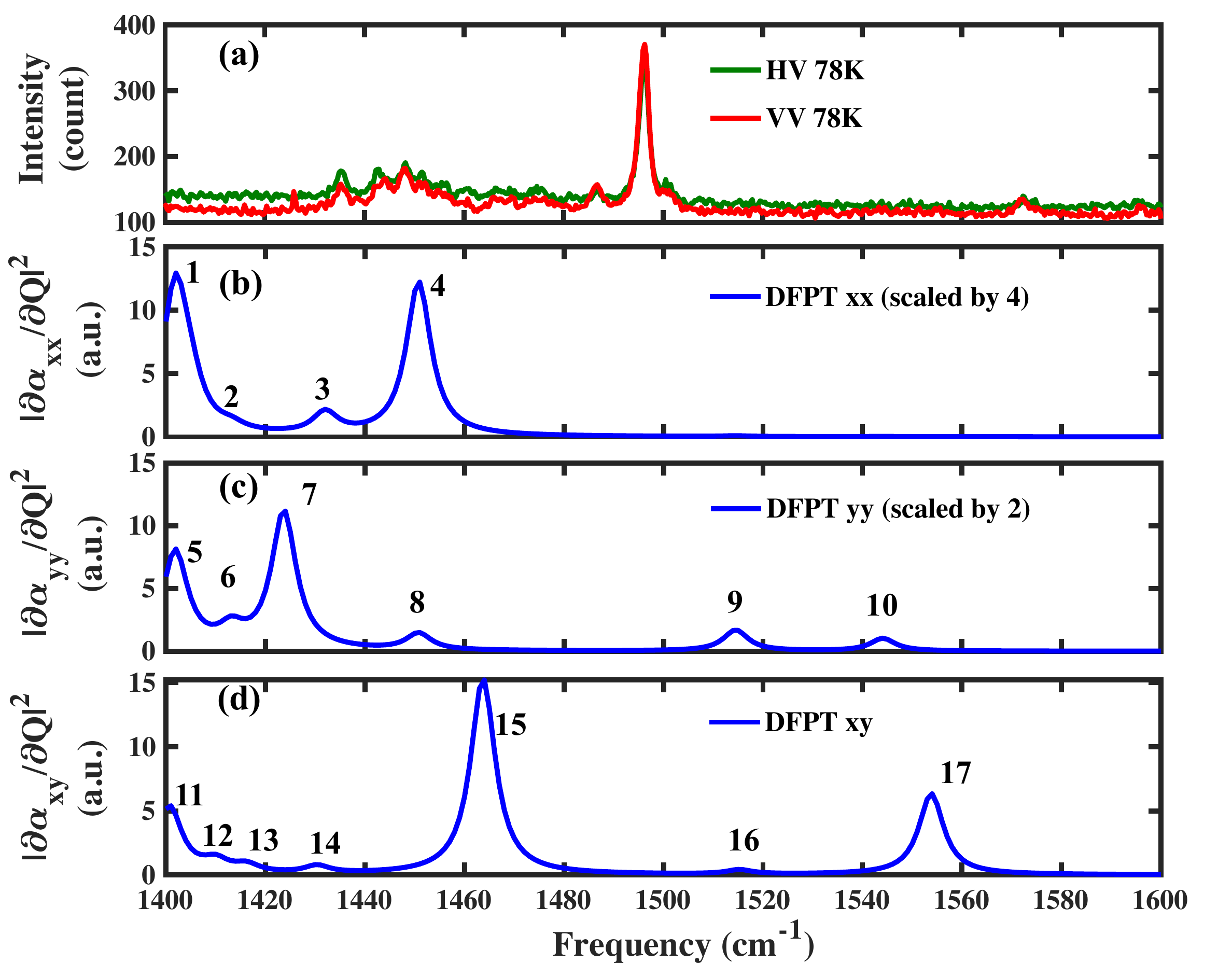}
 \caption{The Raman activities of \haa~between 1400 \cm~and 1600 \cm. (a) Measured Raman intensities in HV (green) and VV (red) polarizations. (b) The $xx$ contribution to the calculated Raman intensity, scaled up by 4 times. (c) The $yy$ contribution to the calculated Raman intensity, scaled up by 2 times. (d) The $xy$ contribution to the calculated Raman intensity. Peak numbers are shown to facilitate references in Table S1 and Figure S1.}
 \label{fig:haa}
 \end{figure}

We carry out the same analysis for the other three materials. For \haa, Figure \ref{fig:haa}(a) shows the experimental Raman spectra, and Figures \ref{fig:haa}(b)(c)(d) show the three contributions to the calculated Raman intensities: $xx$ in (b), corresponding to the matrix element $a$ in Eq. \eqref{eq:ha}, $yy$ in (c), corresponding to the matrix element $b$ in Eq. \eqref{eq:ha}, and $xy$ in (d), corresponding to the matrix element $e$ in Eq. \eqref{eq:ha}. We label each peak in Figures \ref{fig:haa}(b)(c)(d), and show the values of the frequencies and each orientation-dependent contributions to the Raman intensities in Table S1, and the schematic view of the normal-mode eigenvectors in Figure S1. There are again six Raman-active modes localized on the -NH$_3$ group, which are the last three rows in Table S1 and the last six panels in Figure S1. From DFPT calculations, the largest Raman intensities are from a mode at 1463.7 \cm~and another one at 1553.8 \cm, both of which are ammonium bending modes.

\begin{figure}[htbp]
\centering
\includegraphics[width=0.6\columnwidth]{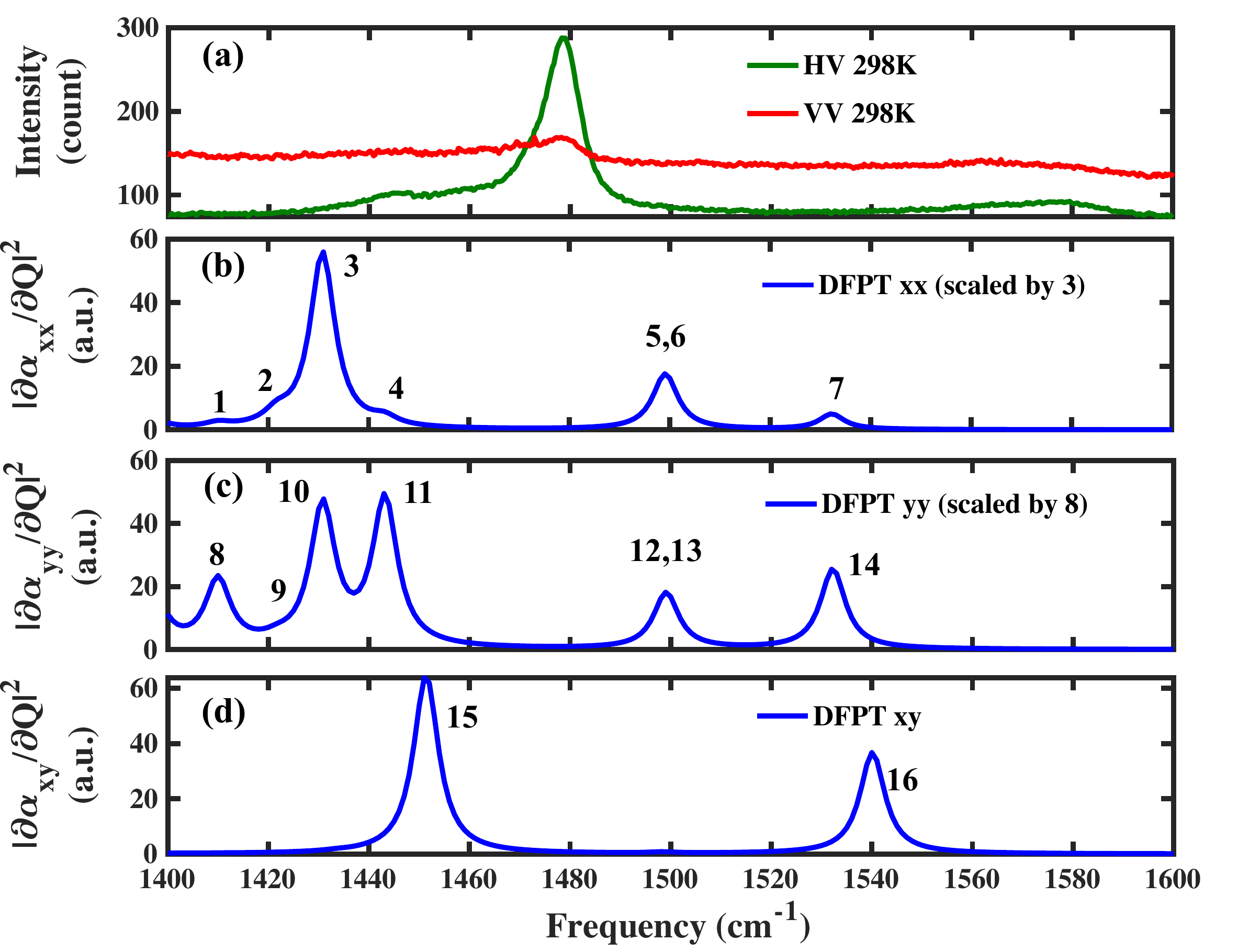}
 \caption{The Raman activities of \ba~between 1400 \cm~and 1600 \cm. (a) Measured Raman intensities in HV (green) and VV (red) polarizations. (b) The $xx$ contribution to the calculated Raman intensity, scaled up by 3 times. (c) The $yy$ contribution to the calculated Raman intensity, scaled up by 8 times. (d) The $xy$ contribution to the calculated Raman intensity. Peak numbers are shown to facilitate references in Table S2 and Figure S2.}
 \label{fig:ba}
 \end{figure}

\begin{figure}[htbp]
\centering
\includegraphics[width=0.6\columnwidth]{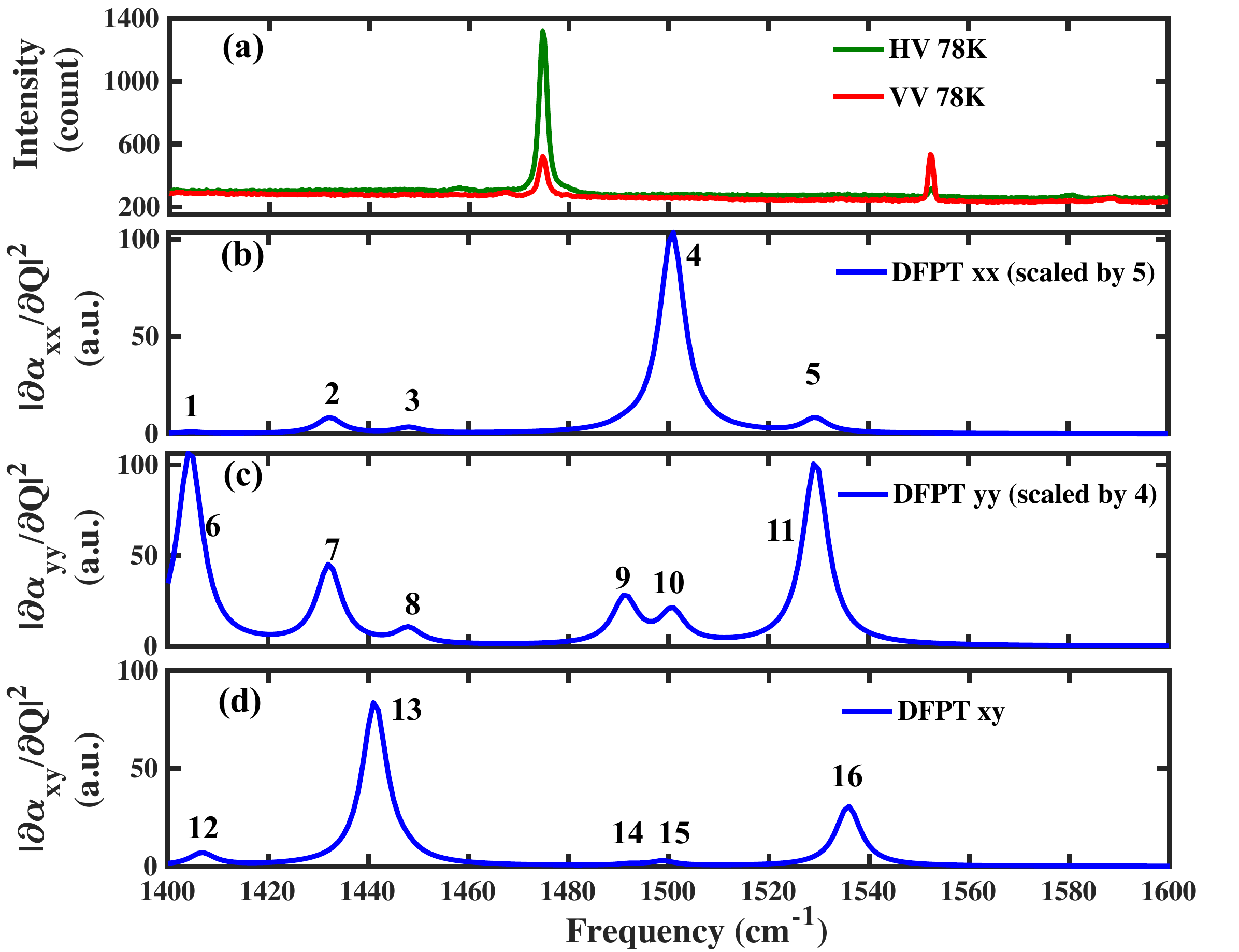}
 \caption{The Raman activities of \bna~between 1400 \cm~and 1600 \cm. (a) Measured Raman intensities in HV (green) and VV (red) polarizations. (b) The $xx$ contribution to the calculated Raman intensity, scaled up by 5 times. (c) The $yy$ contribution to the calculated Raman intensity, scaled up by 4 times. (d) The $xy$ contribution to the calculated Raman intensity. Peak numbers are shown to facilitate references in Table S3 and Figure S3.}
 \label{fig:bna}
 \end{figure}

For \ba~[\bna], Figure \ref{fig:ba} (Figure \ref{fig:bna}) shows the experimental Raman spectra in (a) and the three contributions to the calculated Raman intensities in (b)(c)(d): $xx$ in (b), corresponding to the matrix element $a$ in Eq. \eqref{eq:ba}, $yy$ in (c), corresponding to the matrix element $b$ in Eq. \eqref{eq:ba}, and $xy$ in (d), corresponding to the matrix element $d$ in Eq. \eqref{eq:ba}. We label each peak in Figure \ref{fig:ba} (Figure \ref{fig:bna}), and show the values of the frequencies and each orientation-dependent contributions to the Raman intensities in Table S2 (Table S3), and the schematic view of the normal-mode eigenvectors in Figure S2 (Figure S3). In these two materials, we recognize that the modes within 1400-1600 \cm~range can be divided into groups with 8 modes in each group having similar frequencies, consistent with the 8 molecules in the unit cell. Among the 8 modes in each group, 4 are Raman active, with A$_{\rm g}$, B$_{\rm 1g}$, B$_{\rm 2g}$, and B$_{\rm 3g}$ symmetries, respectively, as can be seen in Tables S2 and S3. There are 12 Raman-active modes localized on the -NH$_3$ group, which are shown in the last three rows of Table S2 for \ba~[the second row and the last two rows of Table S3 for \bna]. The calculated largest Raman intensities are from modes at 1451.4 \cm~and 1540.1 \cm~for \ba~[1441.3 \cm~and 1535.8 \cm~for \bna], all of which are ammonium bending modes.

\subsection{Different Density Functionals and Quantitative Comparison with Experiment} 

In Figure \ref{fig:ha}, we notice the apparent discrepancy (about 30 \cm) between experimental and computational results in the frequencies of the predominant peaks. We attribute this to the intrinsic error of the LDA functional, which is known to underestimate the phonon frequencies\cite{PhysRevB.93.195206}. Unambiguous assignment of experimental peaks requires a correction to the errors made by the underlying density functional used in the DFPT calculations. To this end, we follow and generalize the strategy employed in Ref. \citenum{doi:10.1021/acs.jpcc.5b07432}. The basic assumption is that LDA-based DFPT yields reasonably accurate eigenvectors (normal modes) of the dynamical matrix, but not eigenvalues (phonon frequencies). The remedy is the following: we first compute total energies using a different functional - preferably one that is more accurate than LDA - for a series of geometries along a 1D potential energy curve where the atoms are displaced based on a specific normal mode of interest (which is from LDA-based DFPT), and then calculate the phonon frequencies corresponding to the new functional via a quadratic fitting of the 1D potential energy curve. This frozen-phonon approach is based on the harmonic approximation, and requires that the atomic displacements are small. Before a new functional is used, we use LDA to recalculate the frequency of the normal mode of interest using the frozen-phonon approach, and compare the result against that of the LDA-based DFPT. In the ideal case, we expect that LDA in the frozen-phonon approach yields the same frequency as DFPT for the same mode, and the frequency computed using the more accurate functional in the frozen-phonon approach is then supposed to be a better prediction to the experimental measurement than LDA-based DFPT.

In the frozen-phonon approach, the displacement of an atom $I$ along the spatial direction $\alpha$ ($=x,y,$ or $z$) in the normal mode $\nu$ is \cite{doi:10.1021/acs.jpcc.5b07432,Giustino} 
\begin{equation}
\Delta_{I\alpha,\nu} = \sqrt{\frac{M_0}{M_I}}\mathbf{e}_{I\alpha,\nu}\delta_{\nu},
\label{eq:fpdelta}
\end{equation}
where $\delta_{\nu}$ is the magnitude of the displacement along the normal-mode coordinate and is small to ensure the harmonic approximation, $M_0$ is the reduced mass, $M_I$ is the mass of atom $I$, and $\mathbf{e}_{I\alpha,\nu}$ is the normalized eigenvector of the dynamical matrix for mode $\nu$. After the total energies of a series of displaced structures $E(\delta_{\nu})$ are calculated using the new functional, the corrected frequency for mode $\nu$ can be calculated using
\begin{equation}
\omega_{\nu}^2=\frac{1}{M_0}\frac{d^2E(\delta_{\nu})}{d\delta_{\nu}^2}.
\end{equation}

\begin{figure}[h]
\centering
\includegraphics[width=\textwidth]{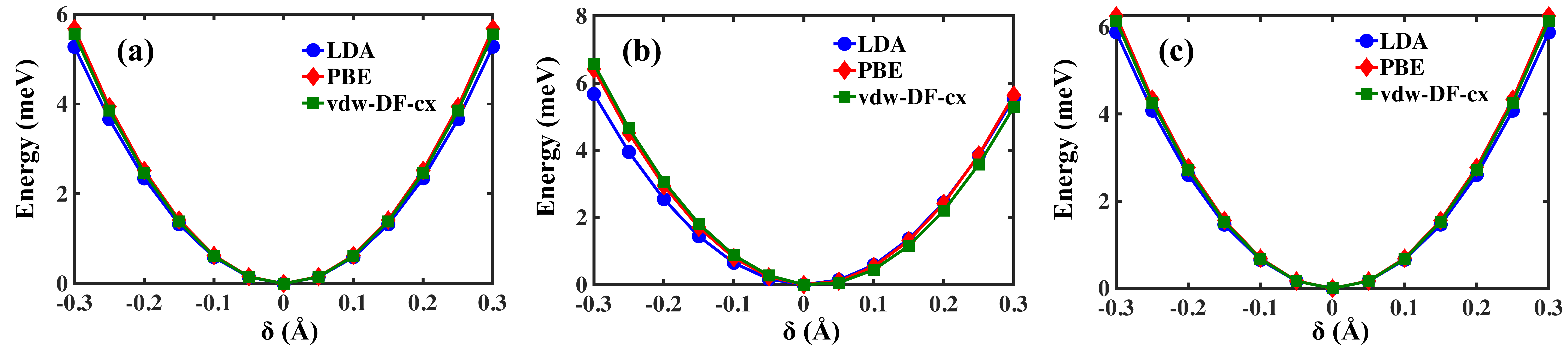}
\caption{1D potential energy curves along the normal-mode coordinates of \ha, for (a) peak 13 in Figure \ref{fig:ha} (1449.8 \cm~using LDA-based DFPT); (b) peaks 4 and 10 in Figure \ref{fig:ha} (1500.6 \cm~using LDA-based DFPT); and (c) peak 15 in Figure \ref{fig:ha} (1536.4 \cm~using LDA-based DFPT). LDA (blue), PBE (red), and vdw-DF-cx (green) functionals are used. Solid lines represent the quadratic fitting used to calculate the phonon frequency from each functional. The energy of the equilibrium structure is set to be zero in all curves.}
\label{fig:fp}
\end{figure}

Figure \ref{fig:fp} shows our results using the frozen-phonon approach, for three representative modes of \ha. We consider the last three groups of modes in Table \ref{tab:ha}, which are the ammonium bending modes. We choose the mode with the higher Raman intensity in each group: 1449.8 \cm~(peak 13 in Figure \ref{fig:ha}), 1500.6 \cm~(peaks 4 and 10 in Figure \ref{fig:ha}), and 1536.4 \cm~(peak 15 in Figure \ref{fig:ha}). In the frozen-phonon approach, we displace the atoms along the normal-mode coordinates up to 0.3 \AA~from the equilibrium structure, at an interval of 0.05 \AA. To be specific, for each mode $\nu$ we examine, the $3N$ ($N$=number of atoms) coordinates of the structure are displaced according to Eq. \eqref{eq:fpdelta}, where $\delta_\nu=0.05,0.1,\cdots,0.3$ \AA. A quadratic form is used to fit the 1D energy curves in Figure \ref{fig:fp}. We first use LDA in the frozen-phonon approach to verify that it could reasonably reproduce the DFPT results (see discussions below). We then proceed to use two other functionals, the Perdew-Burke-Ernzerhof (PBE) \cite{PhysRevLett.77.3865} and the vdw-DF-cx \cite{vdwdfcx} to evaluate the phonon frequencies. Similar calculations are performed for high Raman intensity modes in other materials, and we show the corresponding 1D potential energy curves in the Supporting Information: \haa~in Figure S4, \ba~in Figure S5, and \bna~in Figure S6.

\begin{table}[h]
\begin{tabular}{cccccccc}
\hline\hline
 & Peak & \multirow{2}{*}{Experiment} & DFPT & \multicolumn{4}{c}{Frozen phonon} \\
 & Number & & (LDA) & LDA & PBE & vdw-DF-cx & scaled \\
\hline\hline
\ha & 13 & 1484.7 & 1449.8 & 1448.8 & 1503.7 & 1487.2 & 1488.2 \\
(Figure \ref{fig:ha}) & 4 \& 10 & 1553.5 & 1500.6 & 1495.1 & 1549.1 & 1536.2 & 1541.9 \\
 & 15 & 1562.9 & 1536.4 & 1528.5 & 1577.5 & 1563.0 & 1571.1 \\
\hline
\haa & 15 & 1496.3 & 1463.7 & 1427.5 & 1478.8 & 1461.1 & 1498.2 \\
(Figure \ref{fig:haa}) & 17 & 1572.6 & 1553.8 & 1546.7 & 1594.1 & 1577.4 & 1584.7 \\
\hline
\ba & 15 & 1479.0 & 1451.3 & 1418.7 & 1464.4 & 1447.8 & 1481.1 \\
(Figure \ref{fig:ba}) & 16 & 1579.2 & 1540.1 & 1535.1 & 1580.5 & 1566.1 & 1571.2 \\
\hline
\bna & 13 & 1474.8 & 1441.3 & 1406.8 & 1459.4 & 1442.9 & 1478.3 \\
(Figure \ref{fig:bna}) & 16 & 1552.7 & 1535.8 & 1526.6 & 1577.0 & 1562.4 & 1571.8 \\
\hline\hline
\end{tabular}
\caption{Comparison between selected normal modes from experiments, LDA-based DFPT, and the frozen-phonon approach using different functionals. Peak numbers refer to labelled peaks in relevant figures for each material and correspond to modes with the strongest Raman coefficients in the 1400-1600 \cm~range. Experimental frequencies are the peak positions in panel (a) of each relevant figure. ``Scaled'' means using the ratio of vdw-DF-cx and LDA frequencies in the frozen-phonon approach to scale the LDA-based DFPT frequency. All frequencies are in \cm.}
\label{tab:fp}
\end{table}

Table \ref{tab:fp} summarizes the frozen-phonon results for all materials. Except for peak 15 of \haa, peak 15 of \ba, and peak 13 of \bna, LDA in the frozen-phonon approach does reproduce the phonon frequencies directly calculated using DFPT within 10 \cm, as expected. For such cases, LDA (both within the DFPT and the frozen-phonon formalisms) underestimates the experimental frequencies by about 3\%, while PBE overestimates the experimental frequencies by about 1\%, as expected. \cite{PhysRevB.77.165107} The state-of-the-art van der Waals density functional, vdw-DF-cx \cite{vdwdfcx}, quantitatively agrees with experimental measurements. It is however unclear to us why the situation is different for the aforementioned three modes, where the discrepancy between LDA in the frozen-phonon approach and DFPT can be as large as 40 \cm. We suspect that the shallow potential energy surface of the organic cations in such materials \cite{park2018}, and/or the breakdown of the harmonic approximation for some modes \cite{doi:10.1021/acs.jpcc.0c04573} may be the possible reason for this discrepancy. We defer a thorough investigation of this issue to a future work.

Here, we take a rather practical approach, generalizing Ref. \citenum{doi:10.1021/acs.jpcc.5b07432}. We consider the ratio of frequencies calculated from vdw-DF-cx and LDA in the frozen-phonon approach. This ratio characterizes the difference in the curvatures of the 1D potential energy curves (i.e., interatomic force constants) calculated by the two functionals. We  think that this ratio reflects the difference of these two functionals in describing phonon properties and therefore use this ratio to scale the frequencies calculated from LDA-based DFPT. The results are shown in the last column of Table \ref{tab:fp}. One can see that this strategy yields quantitative agreement between experimental and computational results, enabling us to unambiguously assign experimental peaks with specific normal modes.

Overall, the frozen-phonon approach based on DFPT normal-mode eigenvectors could reliably and efficiently correct intrinsic errors of the underlying functional used in the DFPT calculation, LDA in this case. When an advanced functional is used, such as the vdW-DF-cx for the 2D HOIP here, this approach leads to quantitative agreement between the vibrational frequencies found in experiment and computation. Given the complexity of the 2D HOIP unit cells, such agreement is of paramount significance in accurately assigning each peak in the experimental Raman spectra.

\subsection{Ammonium Bending Modes}

As we see in Figures \ref{fig:ha} and \ref{fig:hamodes}, the ammonium bending modes dominate the Raman intensity in the 1400-1600 \cm~range. For \ha, there are six Raman-active ammonium bending modes in the crystal, which can be divided into three groups, with two modes in each group sharing the same vibrational character of the -NH$_3$ functional group, as we have discussed in Figure \ref{fig:hamodes} and Table \ref{tab:ha}. Given their strong localization on the -NH$_3$ group of the molecule (the displacement of the carbon backbone is negligible in such modes, see Figure \ref{fig:hamodes}), these three groups of bending modes are reminiscent of the three bending modes in an isolated NH$_3$ molecule, where one symmetric bending (A$_1$ symmetry) and two degenerate asymmetric bending (E symmetry) modes are present. The differences between the -NH$_3$ bending modes in the crystal and those in an isolated NH$_3$ molecule are due to two factors: (i) The carbon backbone of the organic cation (HA$^+$) causes a blueshift in the frequencies and breaks the degeneracy of the two asymmetric bending modes found in an isolated NH$_3$ molecule; (ii) the lattice framework of the crystal that consists of lead and halogen atoms further alters the frequencies of the isolated HA$^+$ cation to those of the crystal. We show below that the crystal effect can be well-described by a PbI$_3^-$ fragment placed next to an isolated HA$^+$ cation.

\begin{figure}[t]
\centering
\includegraphics[width=0.7\columnwidth]{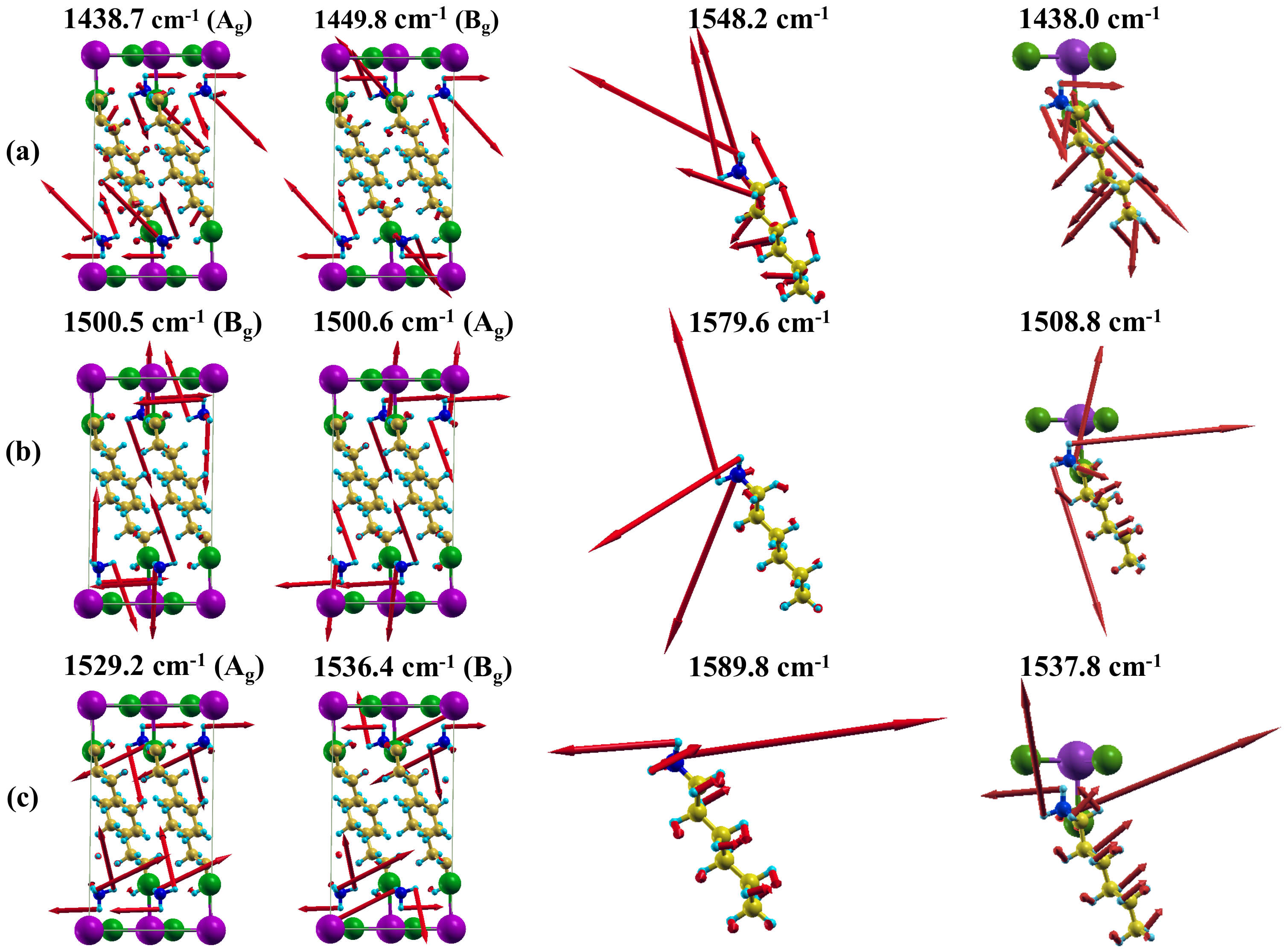}
 \caption{The three groups of Raman-active ammonium bending modes in the \ha~crystal (where there are two modes in each group), compared to the corresponding modes in an isolated HA$^+$ cation and a charge-neutral system that consists of a HA$^+$ cation placed next to a PbI$_3^-$ fragment. All frequencies are from LDA-based DFPT. (a) Symmetric ammonium bending. (b) Asymmetric ammonium bending. (c) Another asymmetric ammonium bending, with different atomic displacement patterns than (b). This figure is rendered using XCrySDen \cite{KOKALJ1999176}.}
 \label{fig:nh3}
 \end{figure}

To this end, we consider an isolated HA$^+$ cation in the gas phase, whose atomic coordinates are the same as one HA$^+$ cation in the optimized \ha~crystal. The HA$^+$ cation is modeled in a large simulation cell of 30 \AA$\times$30 \AA$\times$40 \AA~with a $\mathbf{k}$-mesh of 1$\times$1$\times$1. Moreover, we consider another system, where in addition to the HA$^+$, we place a PbI$_3^-$ fragment nearby, whose atomic coordinates and relative orientations with the HA$^+$ molecule are the same as those in the optimized \ha~crystal. Since lead and iodine atoms form PbI$_6$ octahedra in the crystal, we choose the three iodine atoms forming the ``octant'' in which the -NH$_3$ functional group is located. We have checked that in the crystal, the four HA$^+$ cations are in similar chemical environments in terms of their relative orientations with a neighboring PbI$_3$ fragment. The resulting ``HA$^+$PbI$_3^-$'' unit is charge neutral and is again modeled in a large simulation cell of 30 \AA$\times$30 \AA$\times$40 \AA~with a $\mathbf{k}$-mesh of 1$\times$1$\times$1. We then perform LDA-based DFPT calculations for the isolated HA$^+$ cation and the HA$^+$PbI$_3^-$ system and compare their ammonium bending modes with those in \ha~crystal. Similar analysis is carried out for the other three crystals studied in this work.

Figure \ref{fig:nh3} shows the comparison of normal-mode eigenvectors of the three ammonium bending modes between the \ha~crystal, an isolated HA$^+$ cation, and the HA$^+$PbI$_3^-$ system. Figure \ref{fig:nh3}(a) shows the symmetric bending mode and Figure \ref{fig:nh3}(b)(c) show the two asymmetric bending modes. The cases for other materials are shown in the Supporting Information: \haa~in Figure S7, \ba~in Figure S8, and \bna~in Figure S9. In Figure \ref{fig:nh3}, one can see that for each particular mode, the atomic displacement patterns within the -NH$_3$ functional group are similar among the three systems. We note that strictly speaking, each normal mode consists of vibrational motions of all atoms and the modes shown in Figure \ref{fig:nh3} are only those ``largely localized'' on the -NH$_3$ functional group. The quantitative difference in frequencies between the \ha~crystal and the isolated HA$^+$ molecule can be well explained by the neighboring PbI$_3^-$ fragment. Our finding is consistent with Ref. \citenum{doi:10.1021/jz402589q}, where the PbI$_3^-$ is used to explain the difference in the torsional modes within the 200-400 \cm~range, between an isolated methylammonium cation and the 3D methylammonium lead iodide. 

\begin{figure}[h]
\centering
\includegraphics[width=0.6\columnwidth]{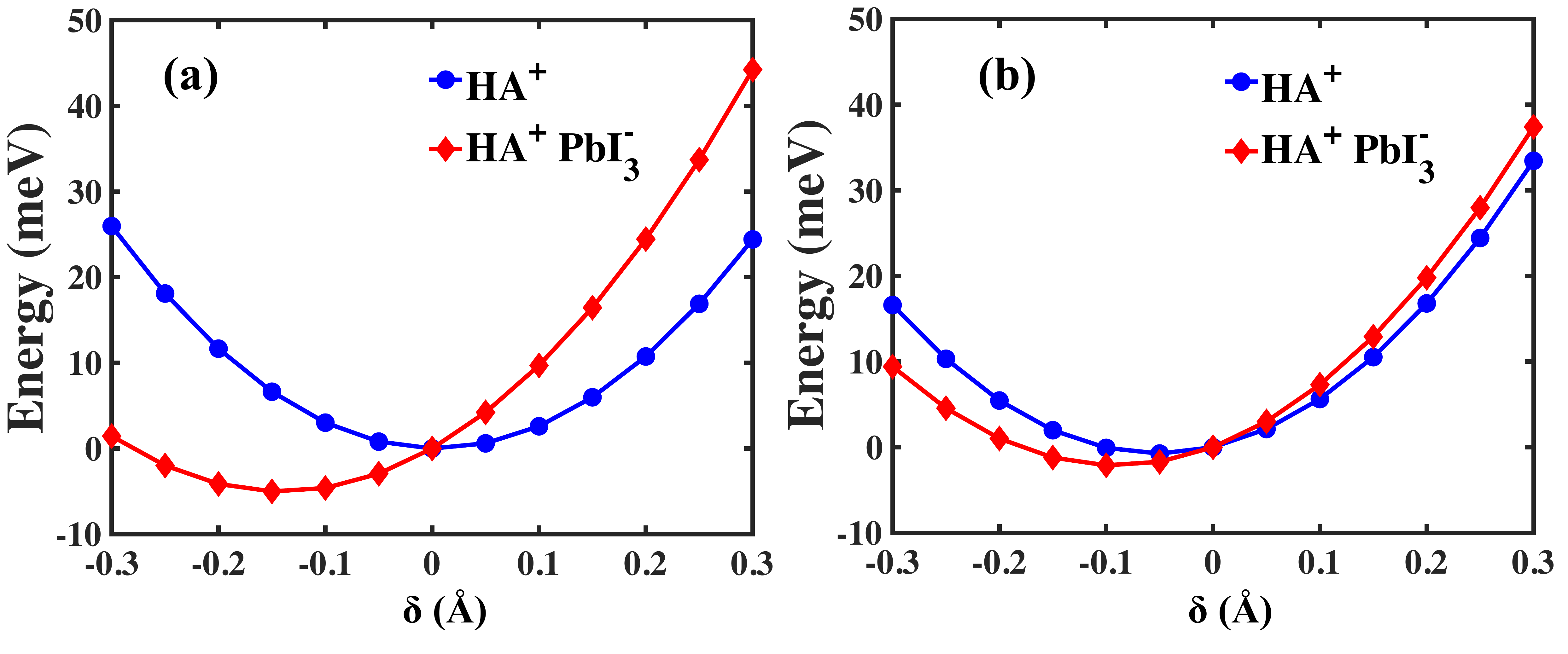}
\caption{1D potential energy curves along the normal-mode coordinates of an isolated HA$^+$, for (a) the asymmetric -NH$_3$ bending mode as shown in Figure \ref{fig:nh3}(b); and (b) another asymmetric -NH$_3$ bending mode as shown in Figure \ref{fig:nh3}(c). Blue dots are calculations of the isolated HA$^+$. Red dots are calculations of the HA$^+$PbI$_3^-$ structure, which is displaced according to the corresponding HA$^+$ normal modes (assuming Pb and I atoms do not move). Solid lines represent the quadratic fitting used to calculate the phonon frequencies. The energy of the non-distorted structure - with the same coordinates as those in the relaxed \ha~crystal - is set to be zero in both systems.}
\label{fig:nh3fp}
\end{figure}

To physically explain the effect of the PbI$_3^-$ fragment in modulating the vibrational frequencies of the isolated HA$^+$, we perform the following analysis. First, we use the aforementioned frozen-phonon approach and the LDA functional to recompute the frequencies of the last two (almost degenerate) HA$^+$ modes in Figure \ref{fig:nh3}. The resulting frequencies are 1582.1 \cm~and 1577.1 \cm, respectively, in good agreement with the LDA-DFPT results (1579.6 \cm~and 1589.8 \cm, see Figure \ref{fig:nh3}). Next, using these HA$^+$ normal modes as the displacement patterns, we compute the frequencies of the HA$^+$PbI$_3^-$ using the frozen-phonon approach, assuming Pb and I atoms do not move. In other words, the HA$^+$ in HA$^+$PbI$_3^-$ is displaced in exactly the same way as the isolated HA$^+$, with a fixed PbI$_3^-$ fragment placed nearby. The resulting frequencies are 1507.9 \cm~and 1526.8 \cm, respectively, again in good agreement with the LDA-DFPT results for HA$^+$PbI$_3^-$ (1508.8 \cm~and 1537.8 \cm, see Figure \ref{fig:nh3}). The 1D potential energy curves for the frozen-phonon calculations of the HA$^+$ and the HA$^+$PbI$_3^-$ are shown in Figure \ref{fig:nh3fp}. Such results indicate that, the presence of PbI$_3^-$ near the HA$^+$ distorts the potential energy surface of the latter, effectively altering the frequencies of the -NH$_3$ localized modes. We note in passing that because the coordinates of both the HA$^+$ and the HA$^+$PbI$_3^-$ are fixed to be those in the relaxed \ha~crystal rather than the individually relaxed species in the gas phase, the energy minima of the two species do not coincide, as we see in Figure \ref{fig:nh3fp}. We have checked that the phonon frequencies of the fully relaxed HA$^+$ and HA$^+$PbI$_3^-$ in the gas phase are very similar (within 10 \cm) to the results reported here.

Additionally, besides PbI$_3^-$, we have also checked explicitly that a neutral PbI$_2$ fragment placed next to the HA$^+$ cation could not quantitatively explain all the frequency differences: it leads to 1439.7 \cm, 1505.7 \cm, and 1567.7 \cm, respectively, for the three modes shown in Figure \ref{fig:nh3}. Although the first two modes are similar to those of the crystal, the frequency of the last asymmetric -NH$_3$ bending mode is apparently off the mark. Moreover, we could rule out the vibrational Stark effect as a possible explanation of the frequency difference between HA$^+$ and the \ha~crystal: the frequency shift under a static electric field of a reasonable intensity [on the order of 0.01 a.u., as determined by the electrostatic potential difference between an isolated HA$^+$ and the \ha~crystal] is only about 10 \cm, much smaller than the 50-100 \cm~that we see in Fig. \ref{fig:nh3}.

\begin{table}[h]
\begin{tabular}{ccrlrlrl}
\hline\hline
& & \multicolumn{2}{c}{Symmetric} & \multicolumn{2}{c}{Asymmetric} & \multicolumn{2}{c}{Asymmetric} \\
& & \multicolumn{2}{c}{-NH$_3$ bending} & \multicolumn{2}{c}{-NH$_3$ bending} & \multicolumn{2}{c}{-NH$_3$ bending} \\
\hline
\multirow{4}{*}\ha & HA$^+$ & 1548.2 & & 1579.6 & & 1589.8 & \\
 & HA$^+$PbI$_3^-$ & 1438.0 & & 1508.8 & & 1537.8 & \\
 & \multirow{2}{*}{crystal} & 1438.7 & A$_{\rm g}$ & 1500.5 & B$_{\rm g}$ & 1529.2 & A$_{\rm g}$ \\
 & & 1449.8 & B$_{\rm g}$ & 1500.6 & A$_{\rm g}$ & 1536.4 & B$_{\rm g}$ \\
\hline
\multirow{4}{*}\haa & HA$^+$ & 1547.0 & & 1578.6 & & 1590.4 & \\
 & HA$^+$PbBr$_3^-$ & 1448.7 & & 1513.3 & & 1544.6 & \\
 & \multirow{2}{*}{crystal} & 1450.9 & A$_{\rm g}$ & 1514.5 & A$_{\rm g}$ & 1544.1 & A$_{\rm g}$ \\
 & & 1463.7 & B$_{\rm g}$ & 1515.2 & B$_{\rm g}$ & 1553.8 & B$_{\rm g}$ \\
\hline
\multirow{6}{*}\ba & BA$^+$ & 1514.8 & & 1590.5 & & 1594.8 & \\
 & BA$^+$PbI$_3^-$ & 1428.8 & & 1508.9 & & 1541.1 & \\
 & \multirow{4}{*}{crystal} & 1442.9 & B$_{\rm 2g}$ & 1498.3 & B$_{\rm 3g}$ & 1532.3 & A$_{\rm g}$ \\
 & & 1443.2 & A$_{\rm g}$ & 1498.4 & B$_{\rm 1g}$ & 1532.4 & B$_{\rm 2g}$ \\
 & & 1451.3 & B$_{\rm 1g}$ & 1499.0 & B$_{\rm 2g}$ & 1540.0 & B$_{\rm 3g}$ \\
 & & 1451.4 & B$_{\rm 3g}$ & 1499.1 & A$_{\rm g}$ & 1540.1 & B$_{\rm 1g}$ \\
\hline
\multirow{6}{*}\bna & BNA$^+$ & 1417.1 & & 1530.8 & & 1569.6 & \\
 & BNA$^+$PbI$_3^-$ & 1408.2 & & 1502.6 & & 1536.8 \\
 & \multirow{4}{*}{crystal} & 1432.1 & B$_{\rm 2g}$ & 1498.7 & B$_{\rm 3g}$ & 1529.2 & B$_{\rm 2g}$ \\
 & & 1432.2 & A$_{\rm g}$ & 1498.8 & B$_{\rm 1g}$ & 1529.4 & A$_{\rm g}$ \\
 & & 1441.1 & B$_{\rm 3g}$ & 1500.7 & B$_{\rm 2g}$ & 1535.8 & B$_{\rm 1g}$ \\
 & & 1441.3 & B$_{\rm 1g}$ & 1500.8 & A$_{\rm g}$ & 1535.9 & B$_{\rm 3g}$ \\
\hline\hline
\end{tabular}
\caption{Comparison of the three -NH$_3$ bending modes in an isolated cation, an isolated cation next to a nearby PbX$_3^-$ (X=Br or I) fragment, and the crystal, for \ha, \haa, \ba, and \bna. All frequencies are calculated from LDA-based DFPT and in \cm.}
\label{tab:nh3}
\end{table}

In Table \ref{tab:nh3}, we summarize the results for all materials. We compare the frequencies of the three ammonium bending modes, for the isolated organic cation, the cation+PbX$_3^-$ (X = Br or I), and the crystal. For all cases, a PbX$_3^-$ placed next to the organic cation successfully explains the difference in frequencies between an isolated organic cation and the 2D HOIP crystal. The only exception is the symmetric -NH$_3$ bending mode in \bna, where a blueshift of the frequency from the isolated cation is found for the crystal, while the use of the PbI$_3^-$ yields a redshift. As we showed in Figure \ref{fig:nh3fp} for \ha, the difference between an isolated cation and the crystal could be physically attributed to the distortion of the potential energy surface of the organic cation by the neighboring PbX$_3^-$, which is local in the crystal. Therefore, these modes listed in Table \ref{tab:nh3} not only feature the strongest Raman intensities in the 1400-1600 \cm~range, but also are sensitive to the local chemical environment around the -NH$_3$ functional group (recall that a PbI$_2$ fragment placed next to the HA$^+$ yields very different frequencies). As a consequence, we think that these ammonium bending modes could serve as a useful experimental probe to the microscopic structure of the lead-halogen octahedra that form the lattice framework of the 2D HOIP crystal. For instance, given the sensitivity of the Raman properties of these ammonium bending modes on the neighboring PbX$_3$ structure, such modes might be able to provide useful information on possible lead or halogen defects in the crystal that give rise to light emission below the materials' optical gaps\cite{Cortecchia2017,Booker2017,Sanni2019a,doi:10.1021/acs.jpclett.9b00743,Kahmann:2020hz,doi:10.1021/acsenergylett.0c01047,Li2020}, which is a point to be scrutinized in future work.

 \section{Conclusion} 
 
In this work, using a combination of Raman spectroscopy in the backscattering geometry and first-principles calculations, we characterized the Raman spectra of four similarly structured two-dimensional hybrid organic-inorganic perovskites, \ha, \haa, \ba, and \bna. We focused on the 1400-1600 \cm~range where the ammonium bending modes dominate the Raman intensity. We first obtained normal-mode eigenvectors from DFPT using LDA, and then performed frozen-phonon calculations using the vdw-DF-cx functional for a series of geometries with atoms displaced along the DFPT normal-mode eigenvectors. We found that this strategy leads to quantitative agreement between experiment and computation, which helped us in unambiguously assigning experimental Raman peaks as normal modes of the crystal. We discussed the relationship between the ammonium bending modes in the periodic crystal and in the isolated organic cation, and found that a PbX$_3^-$ (X = Br or I) fragment placed next to the organic cation could successfully explain the crystal effect on the ammonium bending modes. We conclude that the Raman properties of these ammonium bending modes are useful probes for the microscopic structure of the lead-halogen octahedra in the crystal.

\section{Supporting Information Description}
Tables showing the values of the six orientation-dependent contributions to the Raman intensities of all Raman-active modes of \haa, \ba, and \bna~between 1400 \cm~and 1600 \cm. Figures showing the normal-mode eigenvectors of all Raman-active modes of \haa, \ba, and \bna~between 1400 \cm~and 1600 \cm. Figures showing the frozen-phonon calculations of 1D potential energy curves using LDA, PBE, and vdw-DF-cx, for \haa, \ba, and \bna. Figures comparing the eigenvectors of Raman-active ammonium bending modes between the crystal, an isolated organic cation, and the cation+PbX$_3^-$ system, for \haa, \ba, and \bna.

\begin{acknowledgement}
We thank Marina Filip for insightful discussions of the frozen-phonon approach. A.S.R. and Z.-F.L. acknowledge Wayne State University for generous start-up funds, as well as (partial) support from the American Chemical Society Petroleum Research Fund (60003-DNI6 for A.S.R. and 61117-DNI10 for Z.-F.L.). Computational resources were available through the use of the Center for Nanoscale Materials at Argonne National Laboratory, an Office of Science user facility, which was supported by the U.S. Department of Energy, Office of Science, Office of Basic Energy Sciences, under Contract No. DE-AC02-06CH11357. 
\end{acknowledgement}

\bibliography{Molecular_compare_perovskites}


\end{document}


\newpage







%


\maketitle

\begin{table}[t]
\begin{tabular}{cccccccc}
\hline\hline
Peak number in Figure 5 &  &  \multicolumn{4}{c}{A$_{\rm g}$} &  \multicolumn{2}{c}{B$_{\rm g}$} \\
of the main text & Frequency (cm$^{-1}$) & $xx$ & $yy$ & $zz$ & $xz$ & $xy$ & $yz$ \\
\hline\hline
11   & 1400.7 B$_{\rm g}$                         & -    & -    & -    & -    & 5.28  & 4.51 \\
1, 5 & 1401.9 A$_{\rm g}$                        & 2.84 & 3.91 & 0.68 & 2.31 & -     & -    \\ \hline
1, 5 & 1404.8  A$_{\rm g}$                        & 0.70 & 0.04 & 0.03 & 1.70 & -     & -    \\
-    & 1407.2  B$_{\rm g}$                        & -    & -    & -    & -    & 0.04  & 1.77 \\ \hline
12   & 1410.0 B$_{\rm g}$                          & -    & -    & -    & -    & 0.91  & 9.89 \\
2, 6 & 1413.2 A$_{\rm g}$                        & 0.12 & 0.74 & 0.43 & 0.39 & -     & -    \\ \hline
13   & 1416.2  B$_{\rm g}$                       & -    & -    & -    & -    & 0.62  & 2.79 \\
7    & 1423.7   A$_{\rm g}$                      & 0.00 & 5.51 & 0.51 & 0.02 & -     & -    \\ \hline
14   & 1430.3  B$_{\rm g}$                       & -    & -    & -    & -    & 0.58  & 0.80 \\
3    & 1432.0 A$_{\rm g}$                        & 0.43 & 0.01 & 2.16 & 0.49 & -     & -    \\ \hline
4, 8 & 1450.9 A$_{\rm g}$			 & 3.05 & 0.66 & 0.23 & 0.13 & -     & -    \\
15   & 1463.7 B$_{\rm g}$			 & -    & -    & -    & -    & 15.33 & 0.03 \\ \hline
9    & 1514.5 A$_{\rm g}$				& 0.01 & 0.84 & 0.01 & 1.42 & -     & -    \\
16   &1515.2 B$_{\rm g}$			 & -    & -    & -    & -    & 0.34  & 1.91 \\ \hline
10   & 1544.1 A$_{\rm g}$			 & 0.01 & 0.51 & 0.04 & 0.08 & -     & -    \\
17   & 1553.8 B$_{\rm g}$			 & -    & -    & -    & -    & 6.33  & 0.01  \\ \hline\hline
\end{tabular}
\caption{All Raman active modes of \haa~between 1400 \cm~and 1600 \cm. The corresponding peak numbers in Figure 5 of the main text, and the values of the six orientation-dependent contributions to the Raman intensity are shown next to each mode. All Raman intensities are in atomic unit, i.e., Bohr$^4$/amu, where amu is the atomic mass unit.} 
\label{tab:haa}
\end{table}

\begin{figure}[h!]
\includegraphics[width=\columnwidth]{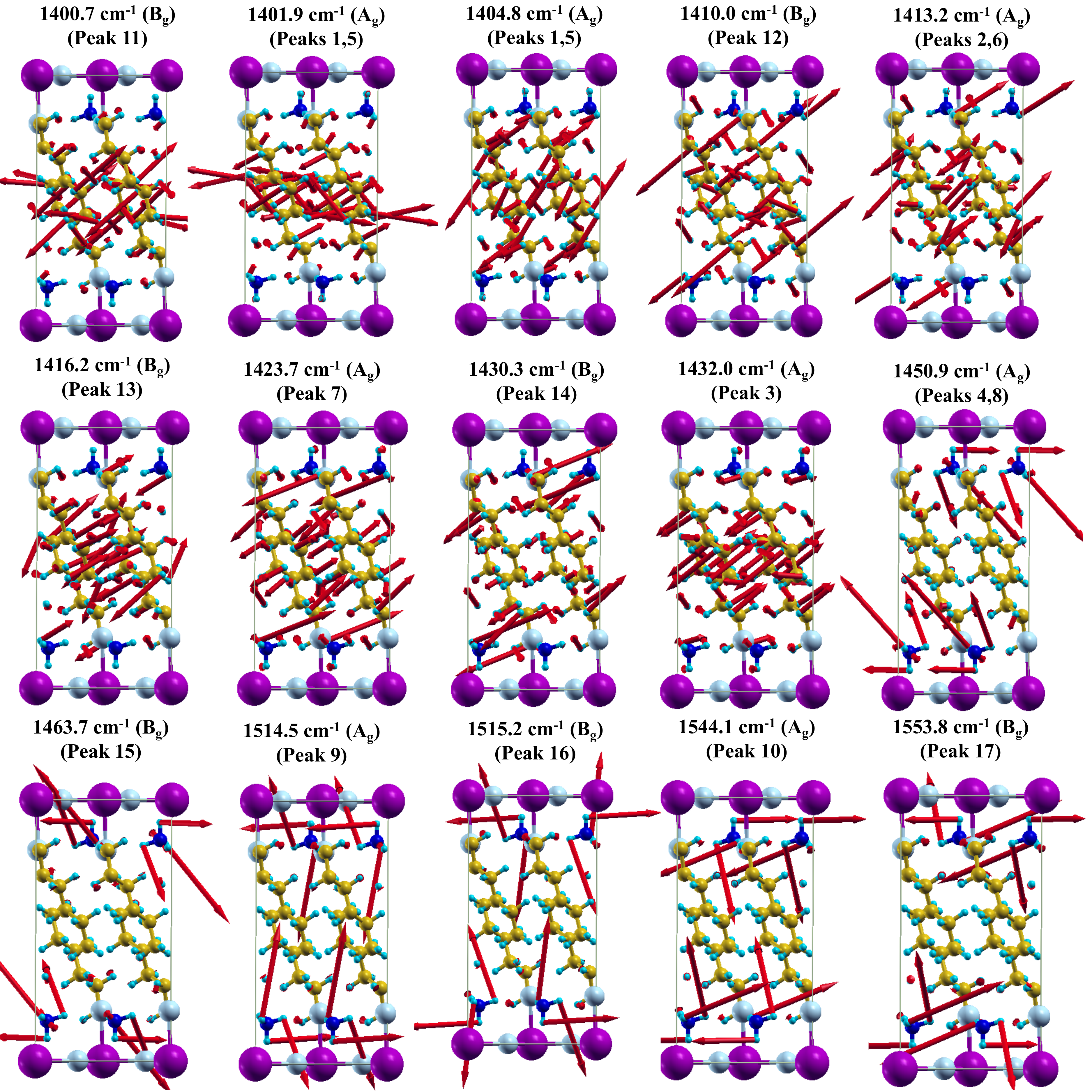}
\caption{A schematic view of the normal mode eigenvectors of all \haa~modes that are included in Figures 5(b)(c)(d) of the main text. The red arrows depict the direction and magnitude of the vibrational movements for each atom. The peak number in Figure 5 of the main text, frequency, and symmetry for each mode are also shown.}
\end{figure}

\begin{table}[t]
\begin{tabular}{cccccccc}
\hline\hline
Peak number in Figure 6 &  & \multicolumn{3}{c}{A$_{\rm g}$} &  B$_{\rm 1g}$ & B$_{\rm 2g}$ & B$_{\rm 3g}$ \\
of the main text  & Frequency (\cm) & $xx$ & $yy$ & $zz$ & $xy$ & $yz$ & $xz$ \\
\hline\hline
-  & 1397.7 A$_{\rm g}$ & 0.78 & 1.73 & 0.07 & - & -     & -      \\
-   & 1402.3 B$_{\rm 1g}$ &  -    & -    & -    & -    & 1.49  & -     \\
-   & 1402.3 B$_{\rm 2g}$ &  -    & -    & -    & 0.11   & -  & -     \\
-   & 1402.7 B$_{\rm 3g}$ &  -    & -    & -    & -    & -  & 3.37 \\ \hline
-    & 1407.6 B$_{\rm 2g}$ &  -    & -    & -    & -    & 6.78  & - \\
-   & 1408.9 B$_{\rm 3g}$ &  -    & -    & -    & -    & - & 9.70     \\
1,8 & 1409.9 A$_{\rm g}$ &  0.48 & 2.66 & 3.67 & - & -     & -    \\
-  & 1410.3 B$_{\rm 1g}$  & -    & -    & -    & 0.06  & - & -     \\ \hline
-   & 1418.9 B$_{\rm 3g}$ &  -    & -    & -    & -    & - & 1.38     \\
-   & 1419.8 B$_{\rm 1g}$ &  -    & -    & -    & 0.01   & -  & -     \\
 -  & 1419.9 B$_{\rm 2g}$ & -    & -    & -    & -    & 2.73 & -  \\
2,9  & 1421.8 A$_{\rm g}$ & 1.28 & 0.18 & 1.42 & - & -     & -    \\ \hline
3,10 & 1430.8 A$_{\rm g}$ & 18.52 & 5.57 & 4.62 & - & -     & -      \\
-   & 1431.3 B$_{\rm 2g}$ &  -    & -    & -    & -    & 0.39  & -     \\
-  & 1433.5 B$_{\rm 1g}$ & -    & -    & -    & -    & -  & 0.83     \\
-   & 1433.6 B$_{\rm 3g}$ &  -    & -    & -    & 0.25   & -  & - \\ \hline
-   & 1442.9 B$_{\rm 2g}$ &  -    & -    & -    & -    & 0.30 & -     \\
4,11 & 1443.2 A$_{\rm g}$ &0.84 & 5.84 & 0.19 & - & -     & -      \\
15  & 1451.3 B$_{\rm 1g}$ & -    & -    & -    & 64.70    & -  & -    \\
-  & 1451.4 B$_{\rm 3g}$ &  -    & -    & -    & -    & - & 0.05  \\ \hline
-   & 1498.3 B$_{\rm 3g}$ &  -    & -    & -    & -    & -  & 5.40 \\
-   & 1498.4 B$_{\rm 1g}$ &  -    & -    & -    & 0.39    & - & -     \\
-   & 1499.0 B$_{\rm 2g}$ &  -    & -    & -    & -    & 4.68 & -     \\
5,6,12,13 & 1499.1 A$_{\rm g}$ & 5.84 & 2.22 & 0.03 & - & -     & -   \\ \hline
7,14 & 1532.3 A$_{\rm g}$ & 1.61 & 3.17 & 0.00 & - & -     & -    \\
-   & 1532.4 B$_{\rm 2g}$ &  -    & -    & -    & -    & 0.08  & -     \\
-   & 1540.0 B$_{\rm 3g}$ &  -    & -    & -    & -    & -  & 0.25    \\
16  & 1540.1 B$_{\rm 1g}$ & -    & -    & -    & 36.70   & - & - \\ \hline\hline
\end{tabular}
\caption{All Raman active modes of \ba~between 1400 \cm~and 1600 \cm. The corresponding peak numbers in Figure 6 of the main text, and the values of the six orientation-dependent contributions to the Raman intensity are shown next to each mode. All Raman intensities are in atomic units.}
\label{tab:ba}
\end{table}

\begin{figure}[h!]
\includegraphics[width=\columnwidth]{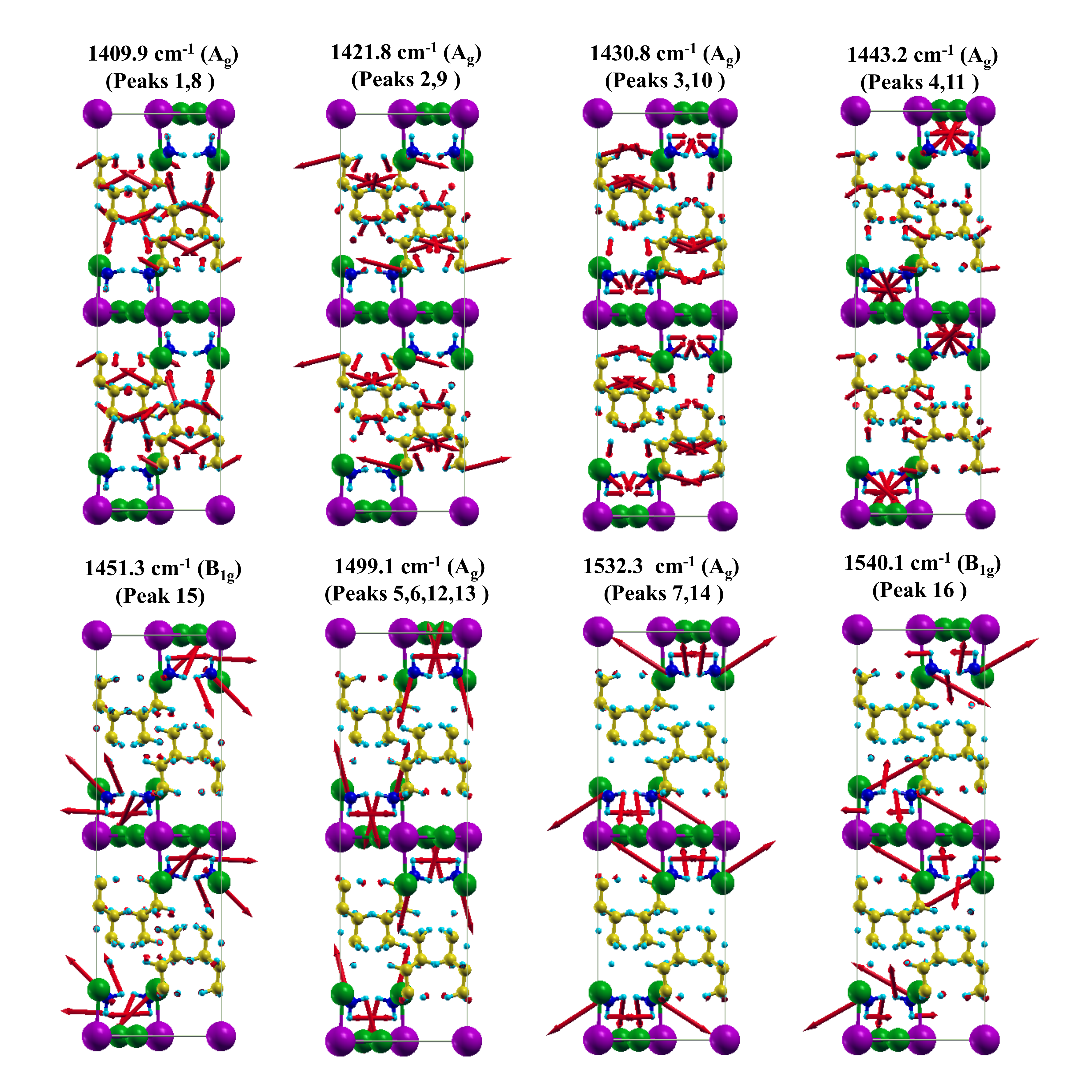}
\caption{A schematic view of the normal mode eigenvectors of all \ba~modes that are included in Figures 6(b)(c)(d) of the main text. The red arrows depict the direction and magnitude of the vibrational movements for each atom. The peak number in Figure 6 of the main text, frequency, and symmetry for each mode are also shown.}
\end{figure}


\begin{table}[t]
\begin{tabular}{cccccccc}
\hline\hline
Peak number in Figure 7 & & \multicolumn{3}{c}{A$_{\rm g}$} & B$_{\rm 1g}$ & B$_{\rm 2g}$ & B$_{\rm 3g}$ \\
of the main text  & Frequency (\cm) & $xx$ & $yy$ & $zz$ & $xy$ & $yz$ & $xz$ \\
\hline\hline
-  & 1404.3 B$_{\rm 2g}$ &  -    & -    & -    & - & 8.40  & -       \\
1,6   & 1404.4 A$_{\rm g}$ & 0.17 & 26.91 & 2.00 & - & -     & -     \\
-   & 1406.6 B$_{\rm 3g}$ &  -    & -    & -    & -  & -  & 0.00     \\
12   & 1406.8 B$_{\rm 1g}$ &  -    & -    & -    & 6.40   & -  & - \\ \hline
-    & 1432.1 B$_{\rm 2g}$ &  -    & -    & -    & -    & 2.42 & - \\
2,7   & 1432.2 A$_{\rm g}$ &  1.63 & 10.89 & 5.19 & - & -     & -      \\
- & 1441.1  B$_{\rm 3g}$ &   -    & -    & -    & -    & -  & 2.87  \\
13 & 1441.3 B$_{\rm 1g}$  & -    & -    & -    & 83.83  & - & -     \\ \hline
-   & 1447.1 B$_{\rm 3g}$ &  -    & -    & -    & -    & - & 0.71    \\
-   & 1447.5 B$_{\rm 1g}$ &  -    & -    & -    & 2.39   & -  & -     \\
 3,8  & 1448.0 A$_{\rm g}$ &0.59 & 2.13 & 4.81 & - & -     & -    \\
- & 1449.2 B$_{\rm 2g}$ &  -    & -    & -    & -    & 3.35 & -    \\ \hline
- & 1490.7 B$_{\rm 2g}$ &  -    & -    & -    & -    & 1.06  & -     \\
9   & 1491.3 A$_{\rm g}$ & 0.29 & 6.51 & 2.47 & - & -     & -       \\
14  & 1492.2 B$_{\rm 1g}$ & -    & -    & -    & 0.81    & -  & -     \\
-   & 1493.0 B$_{\rm 3g}$ &  -    & -    & -    & -  & -  & 0.54 \\ \hline
-   & 1498.7 B$_{\rm 3g}$ &  -    & -    & -    & -    & - & 4.53     \\
15 & 1498.8 B$_{\rm 1g}$ & -    & -    & -    & 2.32   & - & -      \\
-  & 1500.7 B$_{\rm 2g}$ & -    & -    & -    & -    & 4.02  & -    \\
4,10 & 1500.8 A$_{\rm g}$ & 20.78 & 4.47 & 3.40 & - & -     & -  \\ \hline
-   & 1529.2 B$_{\rm 2g}$ &  -    & -    & -    & -    & 0.13  & - \\
5,11  & 1529.4 A$_{\rm g}$ &  1.47 & 25.37 & 3.31 & - & -     & -      \\
16  & 1535.8 B$_{\rm 1g}$ &  -    & -    & -    & 30.67    & - & -     \\
- & 1535.9 B$_{\rm 3g}$ & -    & -    & -    & -    & - & 1.85    \\ \hline\hline
\end{tabular}
\caption{All Raman active modes of \bna~between 1400 \cm~and 1600 \cm. The corresponding peak numbers in Figure 7 of the main text, and the values of the six orientation-dependent contributions to the Raman intensity are shown next to each mode. All Raman intensities are in atomic units.}
\label{tab:bna}
\end{table}

\begin{figure}[h!]
\includegraphics[width=0.9\columnwidth]{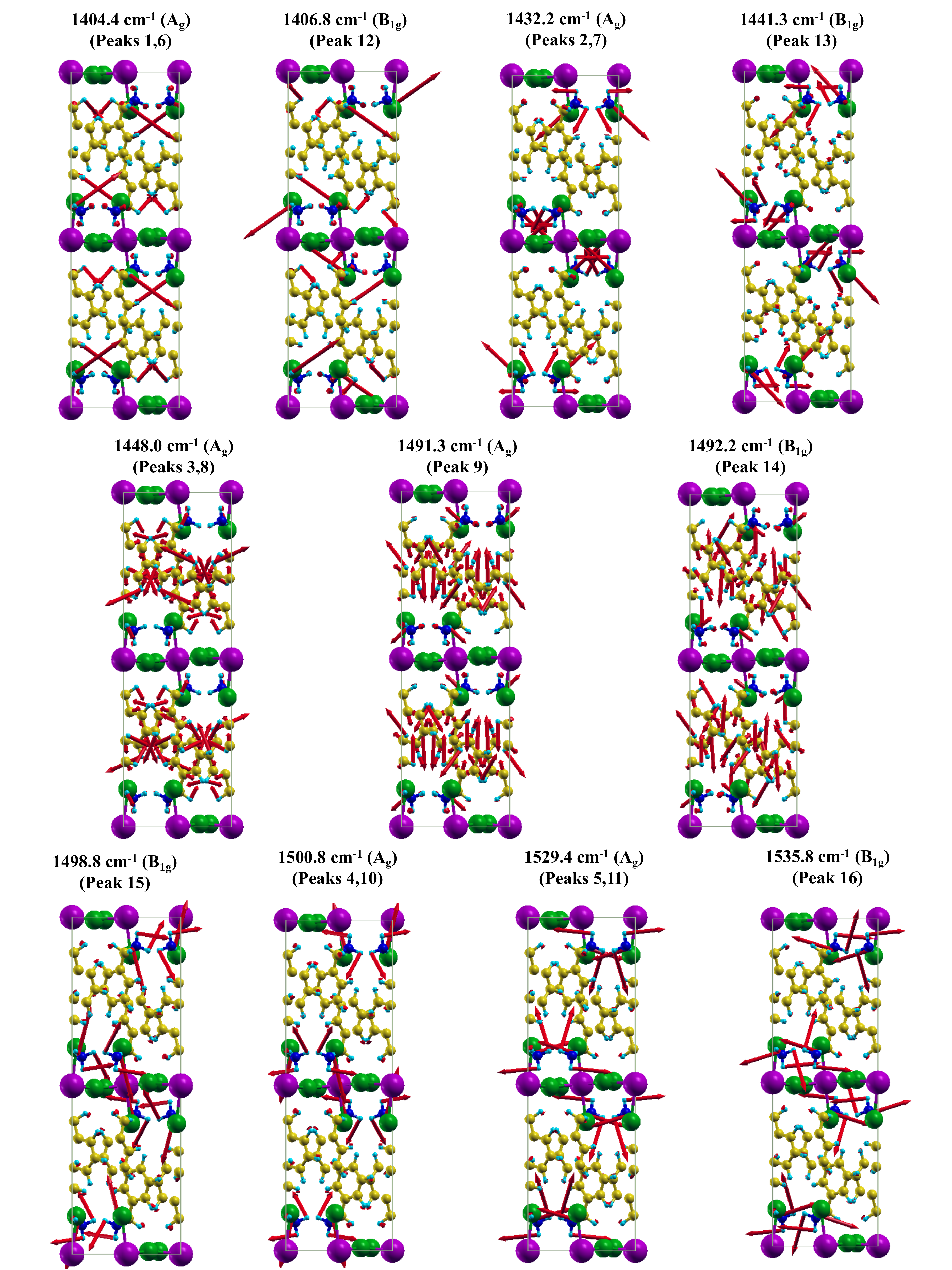}
\caption{A schematic view of the normal mode eigenvectors of all \bna~modes that are included in Figures 7(b)(c)(d) of the main text. The red arrows depict the direction and magnitude of the vibrational movements for each atom. The peak number in Figure 7 of the main text, frequency, and symmetry for each mode are also shown.}
\end{figure}


\begin{figure}[h!]
\includegraphics[width=0.7\columnwidth]{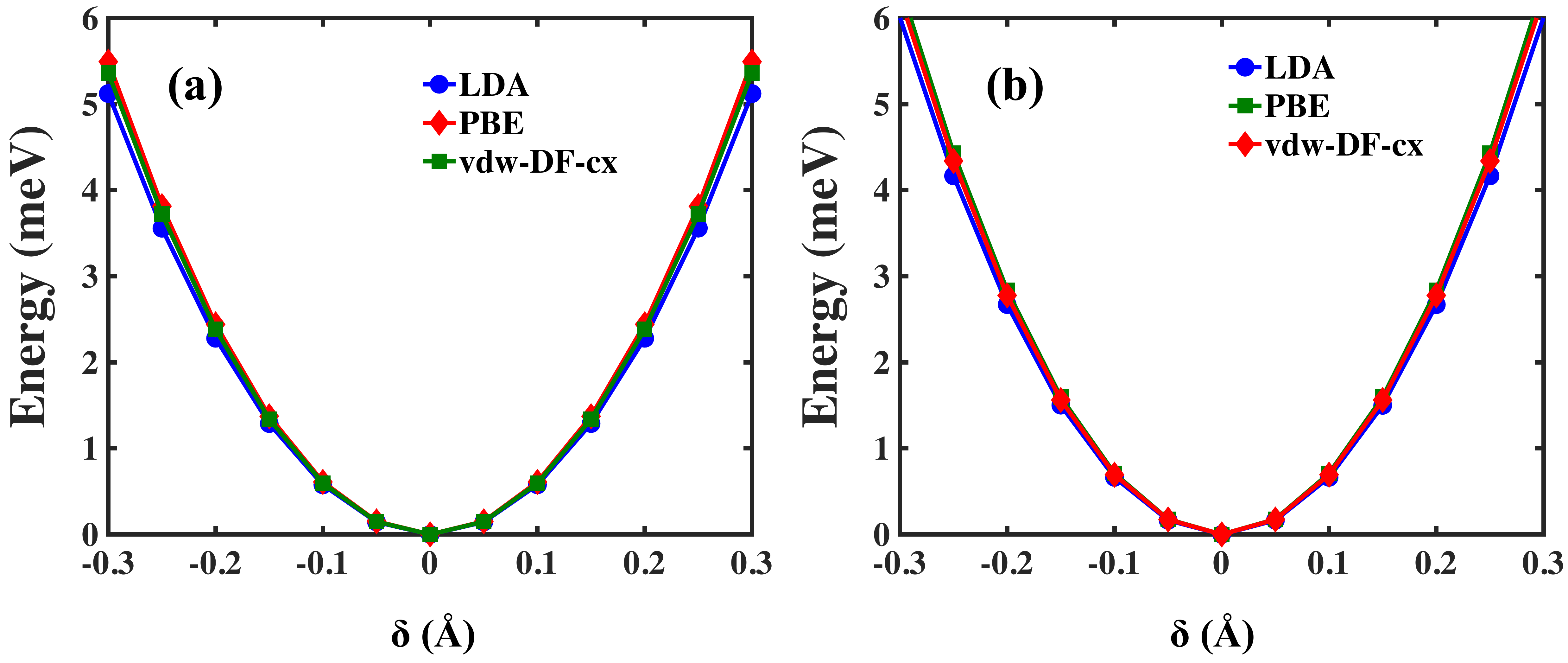}
\caption{One-dimensional potential energy curves along the normal-mode coordinates of \haa, for (a) peak 15 and (b) peak 17 in Figure 5 of the main text. Solid lines represent the quadratic fitting used to calculate the phonon frequency from each functional. }
\end{figure}

\begin{figure}[h!]
\includegraphics[width=0.7\columnwidth]{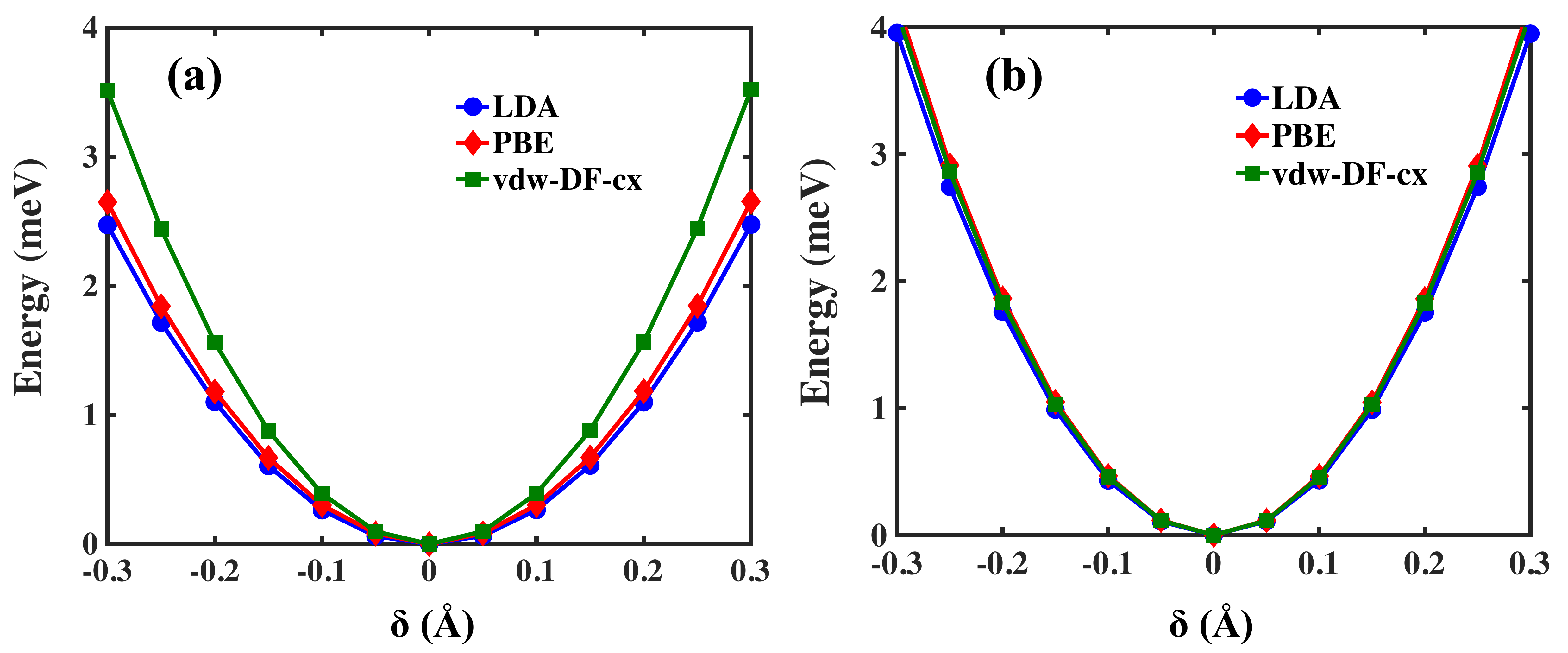}
\caption{One-dimensional potential energy curves along the normal-mode coordinates of \ba, for (a) peak 15 and (b) peak 16 in Figure 6 of the main text. Solid lines represent the quadratic fitting used to calculate the phonon frequency from each functional. }
\end{figure}

\begin{figure}[h!]
\includegraphics[width=0.7\columnwidth]{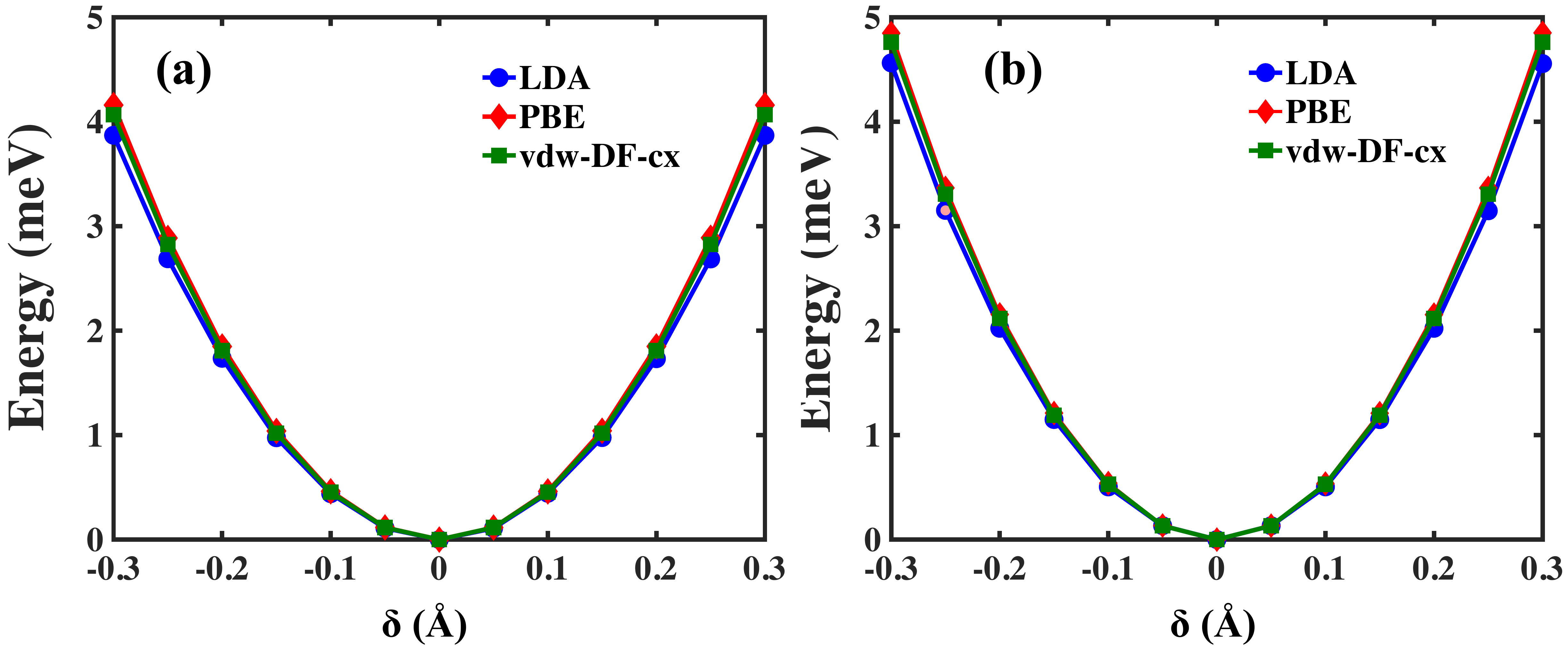}
\caption{One-dimensional potential energy curves along the normal-mode coordinates of \bna, for (a) peak 13 and (b) peak 16 in Figure 7 of the main text. Solid lines represent the quadratic fitting used to calculate the phonon frequency from each functional. }
\end{figure}


\begin{figure}[h!]
\includegraphics[width=\columnwidth]{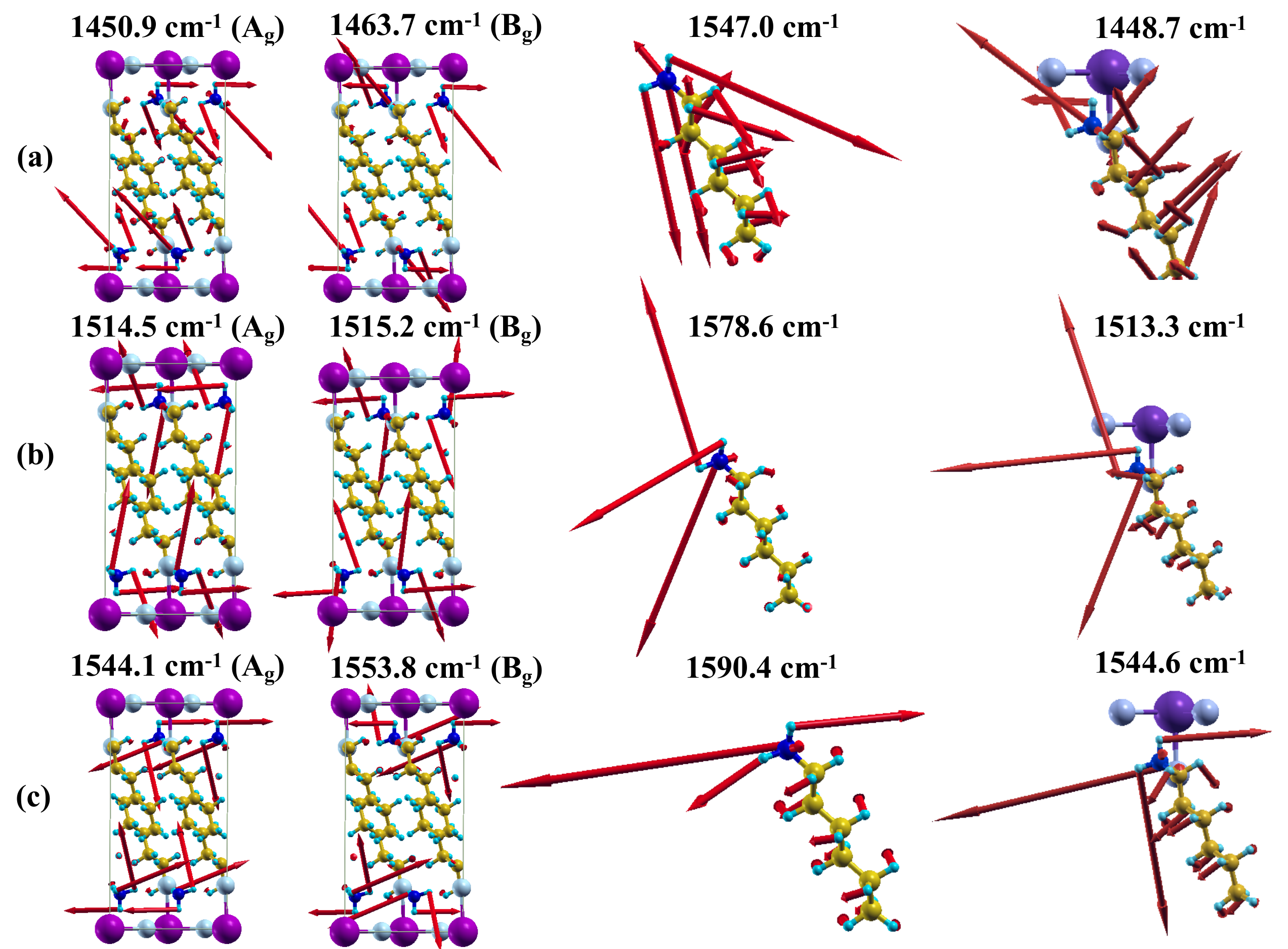}
\caption{The three groups of Raman-active ammonium bending modes in the \haa~crystal (where there are two modes in each group), compared to the corresponding modes in an isolated HA$^+$ cation and a charge-neutral system that consists of a HA$^+$ cation placed next to a PbBr$_3^-$ fragment. All frequencies are from LDA-based DFPT. (a) Symmetric ammonium bending. (b) Asymmetric ammonium bending. (c) Another asymmetric ammonium bending, with different atomic displacement patterns than (b).}
\end{figure}

\begin{figure}[h!]
\includegraphics[width=\columnwidth]{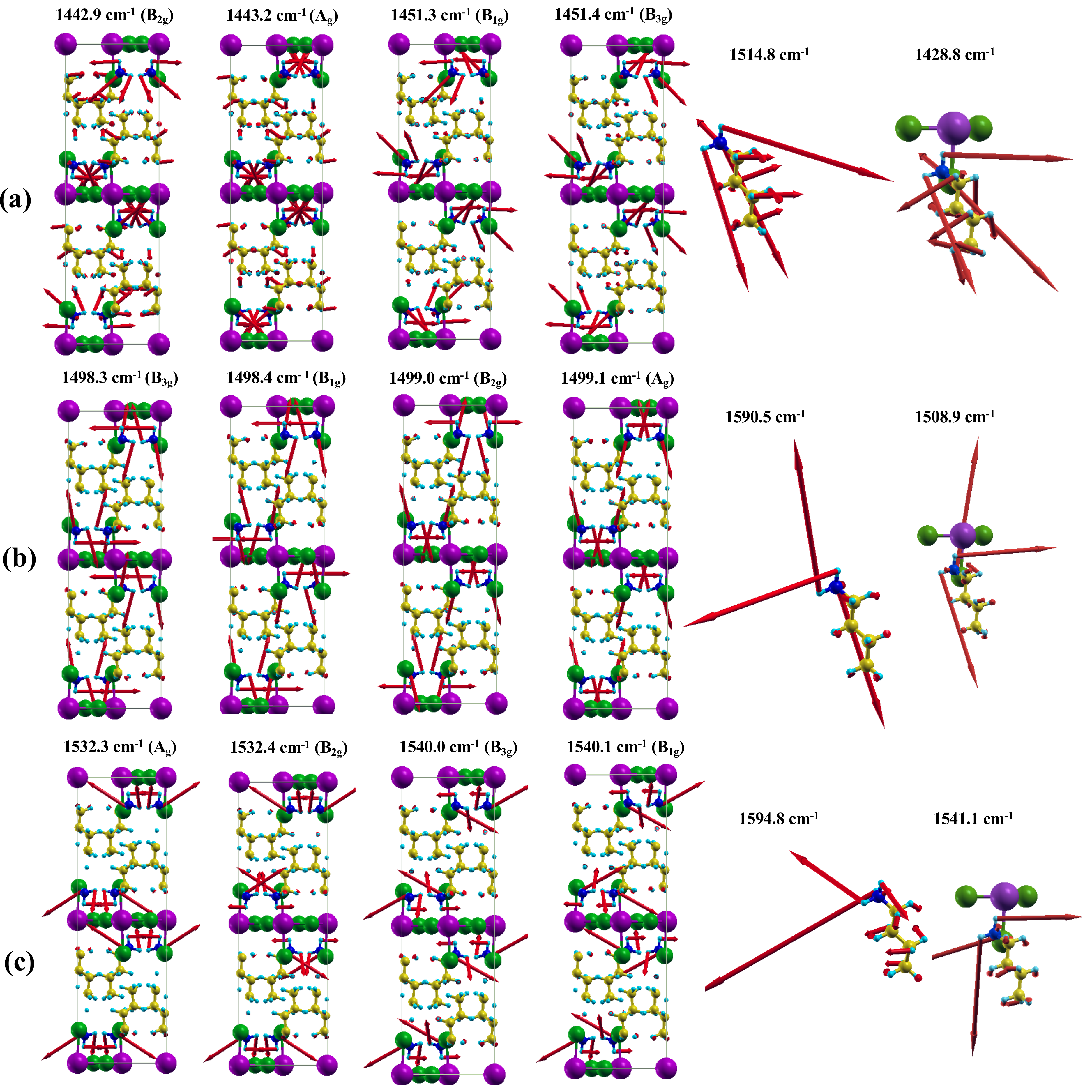}
\caption{The three groups of Raman-active ammonium bending modes in the \ba~crystal (where there are four modes in each group), compared to the corresponding modes in an isolated BA$^+$ cation and a charge-neutral system that consists of a BA$^+$ cation placed next to a PbI$_3^-$ fragment. All frequencies are from LDA-based DFPT. (a) Symmetric ammonium bending. (b) Asymmetric ammonium bending. (c) Another asymmetric ammonium bending, with different atomic displacement patterns than (b).}
\end{figure}

\begin{figure}[h!]
\includegraphics[width=\columnwidth]{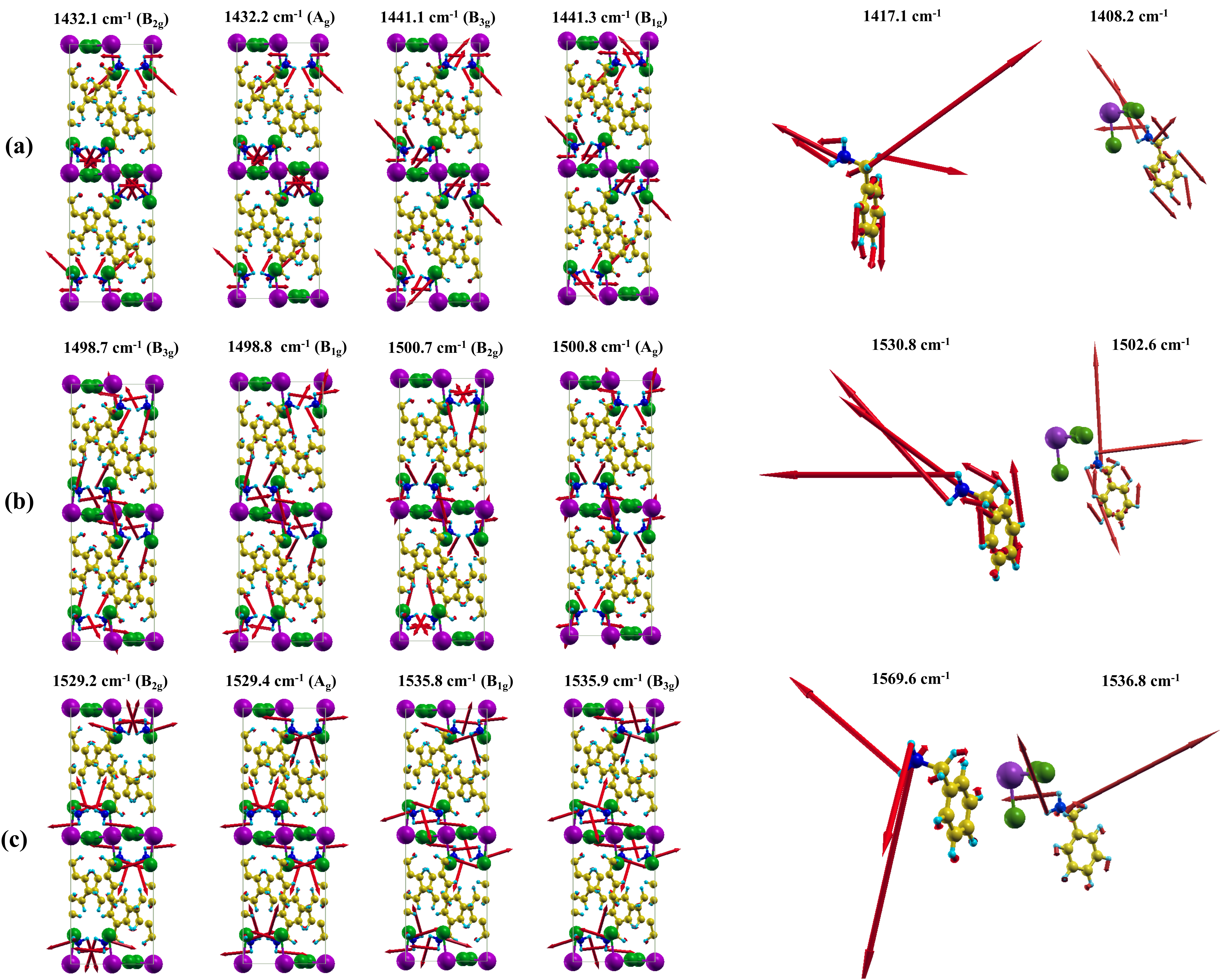}
\caption{The three groups of Raman-active ammonium bending modes in the \bna~crystal (where there are four modes in each group), compared to the corresponding modes in an isolated BNA$^+$ cation and a charge-neutral system that consists of a BNA$^+$ cation placed next to a PbI$_3^-$ fragment. All frequencies are from LDA-based DFPT. (a) Symmetric ammonium bending. (b) Asymmetric ammonium bending. (c) Another asymmetric ammonium bending, with different atomic displacement patterns than (b).}
\end{figure}